%% file: main.tex
\documentclass{article}
\usepackage{arxiv}

\usepackage{amsthm,amsmath}
\usepackage{subcaption}
\usepackage{bbm}
\usepackage{blindtext}
\usepackage{placeins}
\usepackage{booktabs}       
\usepackage{amsfonts}       
\usepackage{microtype}      

\usepackage[numbers]{natbib}
\usepackage{fancyhdr}       
\usepackage{graphicx}
\usepackage{blindtext}
\usepackage{placeins}
\usepackage[hypcap=false]{caption}

\usepackage{bbm}
\usepackage{subcaption}
\graphicspath{{media/}}     
\usepackage{xcolor}

\usepackage[utf8]{inputenc} 

\usepackage{url}            
\usepackage[colorlinks = true,
            linkcolor = blue,
            urlcolor  = blue,
            citecolor = blue]{hyperref}       

\title{Where you go is who you are - A study on machine learning based semantic privacy attacks}

\author{Nina Wiedemann${}^{1, *}$, Ourania Kounadi${}^{2}$, Martin Raubal${}^{1}$, Krzysztof Janowicz${}^{2}$}

\begin{document}

\maketitle
\begin{center}
${}^*$ Corresponding author: nwiedemann@ethz.ch \\
${}^1$ Institute of Cartography and Geoinformation, ETH Zurich, Zurich, Switzerland \\
${}^2$ Department of Geography and Regional Research, University of Vienna, Vienna, Austria \\
\end{center}
\vspace{2em}


\vspace{4em}

\begin{abstract} 
Concerns about data privacy are omnipresent, given the increasing usage of digital applications
and their underlying business model that includes selling user data. Location data is particularly
sensitive since they allow us to infer activity patterns and interests of users, e.g., by categorizing visited locations based on nearby points of interest (POI). On top of that, machine learning methods provide new powerful tools to interpret big data. In light of these considerations, we raise the following question: What is the actual risk that realistic, machine learning based privacy attacks can obtain meaningful semantic information from raw location data, subject to inaccuracies in the data? In response, we present a systematic analysis of two attack scenarios, namely location categorization and user profiling. 
Experiments on the Foursquare dataset and tracking data demonstrate the potential for abuse of high-quality spatial information, leading to a significant privacy loss even with location inaccuracy of up to 200m. With location obfuscation of more than 1 km, spatial information hardly adds any value, but a high privacy risk solely from temporal information remains. 
The availability of public context data such as POIs plays a key role in inference based on spatial information. 
Our findings point out the risks of ever-growing databases of tracking data and spatial context data, which policymakers should consider for privacy regulations, and which could guide individuals in their personal location protection measures.
\end{abstract}


\section*{Introduction}

In the age of big data, an unprecedented amount of information about individuals is publicly available. Not only the information from social media profiles can be exploited to gain rich insights into the private life of individuals, but also data that is collected by applications on-the-fly. Collecting and selling such data has become a business model of commercial consumer data brokers, who distribute individual data of users, oftentimes without their awareness~\citep{crain2018limits}. 
A particularly popular source is location data, as the whereabouts of people allow rich insights into their daily activities~\citep{huang2018location, banerjee2019geosurveillance, duckham2006location, nelson2022accelerating}, for example, for the purpose of profiling. Even though awareness for (location) privacy has increased in recent years~\citep{alrayes2014no}, this is oftentimes not reflected in user behavior, which has been termed the ``privacy paradox''~\citep{potzsch2008privacy, barth2017privacy}. Only gradually, companies are reacting to imposed privacy regulations and the efforts of privacy advocates' groups~\citep{georgiadou2019location}. For example, Apple\textsuperscript{TM} is giving back control over data sharing decisions in the iPhone\textsuperscript{TM}, including location data\footnote{
\url{https://support.apple.com/guide/iphone/control-the-location-information-you-share-iph3dd5f9be/ios}}, and Strava\textsuperscript{TM} offers to restrict track-visibility in their app for recording physical activities.\footnote{\url{https://support.strava.com/hc/en-us/articles/115000173384-Edit-Map-Visibility}} 

The simplest way to protect location data is a form of masking or obfuscation of the exact geographic coordinates~\citep{kounadi_privacy_2018}; i.e., deliberately reducing the data quality~\citep{hutchison_formal_2005}. 
While hiding the exact location may provide some anonymity, the risk of unwanted \textit{semantic} inference from the raw location data remains. For example, if a user is detected in a busy city district at night, it is very likely that the user is in a bar or club. This type of inference was recently termed ''semantic privacy attack''~\citep{tu_protecting_2019}, in contrast to previous work on location privacy that was mainly concerned with user \textit{re-identification} attacks~\citep{de_montjoye_unique_2013, rossi_spatio-temporal_2015, manousakas_quantifying_2018, martin2022influence, krumm_inference_2007}. 


In this work, we define and analyze a special type of semantic privacy attack that is motivated by the real-life problem that brokers obtain location data and sell them as valuable information about user behavior, for example, for targeted advertisement or for insurance policy offer. Here, we disregard how an adversary would \textit{obtain} location data but instead focus on the question of how he would \textit{derive meaningful user profiles} from the raw location data of a single user. 
We argue that a smart attacker would tackle this problem by utilizing spatial and temporal information for categorizing the locations that a user has visited, drawing from methods developed in reverse geocoding~\citep{kounadi2013accuracy, chen2021identifying, efstathiades2015identification, al_hasan_haldar_location_2019, liu_unsupervised_2016, pontes2012we}, activity categorization~\citep{penha2017activity, xiao_detecting_2016, shen_process_2013, montini2014trip, cui_forecasting_2018, falcone_what_2014} and place labeling~\citep{ye_semantic_2011, jenson_mining_2017, do_places_2014} research.
For example, if the location data indicates a two-hour stay in a place with many bars nearby,  
the attacker may derive that the activity falls into the category ''Nightlife''. 
In a second step, the attacker could aggregate the (predicted) categories of all locations that a user visited into a location-based user profile. For example, the profile is 60\% ''Dining'', 30\%  ''Retail'', and 10\% ''Nightlife''. 
In short, we consider the following two semantic attack scenarios:

\begin{itemize}
    \item[] \textbf{Task 1}: Given a location visit defined by geographic coordinates and a visitation time, the attacker aims to assign the place to the correct category. 
    \item[] \textbf{Task 2}: Given the location visitation pattern of a user, the attacker aims to derive a user profile, defined as the visitation frequencies to each of the location categories.
\end{itemize}
To the best of our knowledge, this type of location-based user profiling has not been regarded as a privacy attack, and similar definitions for user profiles are mainly found in literature on recommender systems~\cite{xie2016learning, ying2010mining}. 
Note that if these tasks are feasible, the attacker would not only know about activity frequencies but also about when and where each type of activity is preferably carried out. The input data of the attacker is assumed to consist only of geographic coordinates and timestamps. Such data could stem from GNSS tracking data, from Call-Detail-Records~\citep{yuan2016analyzing}, or other forms of movement data. 

According to \citep{kesler_geoprivacy_2018}, ''an individual’s level of geoprivacy cannot be reliably assessed because it is impossible to know what auxiliary information a third party may have access to.” (p. 11). However, one can attempt to quantify the level of privacy by simulating realistic scenarios and measuring the accuracy of the attacker~\cite{shokri_quantifying_2011, shokri2015quantifying}. By realistic, we mean that an attacker tries to enrich the raw data with as much information as possible and employs sophisticated algorithms to analyze patterns in such information. 
We believe that there is a lack of work analyzing 1) which spatial and temporal information may be exploited, 2) how the data quality, as well as the level of intended inaccuracy due to location protection measures, affects an attacker’s accuracy, and 3) what is the relation to the density and quality of spatial context data, e.g., public POIs. 
We, therefore, evaluate the effectiveness of machine learning based semantic privacy attacks in different scenarios with respect to the information available to the attacker and, similar to~\cite{gao2019exploring}, varying the data accuracy by means of random perturbations of the location. %

\section*{Related work}
\subsection*{Reverse geocoding and activity categorization}

Many studies utilize a well-known dataset of location check-ins from the Location-based Social Network (LBSN) Foursquare, which is very suitable due to its size, its detailed POI categorization taxonomy, and the availability of user-wise check-in data. The POIs and visitation patterns were analyzed for recommender system applications~\citep{ye2011exploiting}, for deriving interpretable latent representations of venues~\citep{an_enabling_2022} or to infer urban land-use via clustering of POI data~\citep{gao2017extracting}. \citet{yang2014modeling} train models on the Foursquare dataset to infer spatio-temporal activity preferences of users for the purpose of place recommendation. In this work, we take a machine learning viewpoint and regard the Foursquare data as a labeled dataset that is suitable to model the real-life scenario where an attacker aims to categorize the locations of an \emph{unseen} user.\\ 
However, it was shown that not only spatial but also temporal information about location visits could be exploited to infer location categories~\citep{mckenzie2015also}. This has been reported implicitly in other work, for example, \citet{do_places_2014} regard the problem of automatic place labeling into 10 categories, leveraging visitation patterns, e.g., temporal features (start and end time or duration) and visitation frequency from smartphone data. \citet{mckenzie2016geo} connect this observation to geoprivacy research by showing that temporal information or texts from social media posts can be exploited for inference about user locations by matching their semantic signatures~\citep{janowicz2012observation, mckenzie2015poi}. While our study is on location categorization and user profiling, in contrast to user localization, their study inspired us to include temporal features in the attack scenario and to contrast their effect on the attacker's success to the one due to spatial information.\\
Furthermore, work on user profiling from location data (our second attack task) can mainly be found in the literature on recommender systems, which is surveyed in \citep{bao2015recommendations}. The POI embedding of users can be viewed as their location profile, for example, with graph-based embeddings \citep{xie2016learning}. 
\citet{ying2010mining} compare users by their ''semantic trajectory'', defined as the categories of sequentially visited places. We follow their approach but disregard the order of places.

\subsection*{Location privacy research}

Privacy risks and potential privacy preservation techniques were studied extensively in the past years~\citep{ram2018privacy}. 
In location privacy research, it was found that a few track points are sufficient to uniquely identify users~\citep{de_montjoye_unique_2013, rossi_spatio-temporal_2015, golle2009anonymity}, that it is possible to track people just by the speed and starting location~\citep{gao2014elastic} or by accelerometer readings~\citep{han2012accomplice}, and that even topological representations of movement data without coordinates can be exploited to match users~\citep{manousakas_quantifying_2018, martin2022influence}. A common aim of many works is to maintain the performance of a location-based service while providing privacy guarantees; i.e., to optimize the privacy-utility trade-off~\citep{sreekumar2019optimal, cerf2017pulp}. 
Various frameworks for protecting sensitive location data were proposed~\citep{hutchison_formal_2005, mckenzie_privyto_nodate, charleux2020true, seidl2016privacy, miranda2023sok}, oftentimes based on k-anonymity~\citep{sweeney2002k, gruteser2003anonymous, gurung2014traffic} or $\epsilon$-differential privacy~\citep{haydari_adaptive_2021, andres2013geo, dwork2008differential, jain2018differential}. For an overview of possible privacy attacks on location data and protection methods we refer to the reviews by \citet{kounadi_privacy_2018} and \citet{wernke2014classification}.\\
This work instead analyzes privacy attacks that aim to reveal personal information, i.e., interests and behavioural patterns. Related work in this direction, for example, investigates to what extent demographics (e.g., age or gender) and visited POIs can be derived from location traces~\citep{li2016privacy}. 
\citet{crandall2010inferring} and \citet{olteanu2016quantifying} analyze co-location events and the risk to infer social ties. \citet{tu_protecting_2019} recently termed the inference of private semantic information from movement trajectories as a ''semantic'' privacy attack, and they specifically regard contextual POI data as semantics. 
We build up on their definition and consider attacks that aim to infer POI categories. \citet{tu_protecting_2019} propose l-diversity and t-closeness measures to protect trajectories from semantic inference. However, these approaches rely on trusted third-party (TTP) services that mask the data of multiple users and update their data iteratively in online applications~\citep{pei2007maintaining, khan2022privacy}. Omitting the dependence on a TTP is possible, for example, with simple location obfuscation methods, i.e., adding random noise to coordinates or methodologically translating geographic coordinates in space~\citep{hutchison_formal_2005, andres2013geo}.  \citet{zhang2018context} and \citet{gotz2012maskit} further propose context-aware masking techniques that are applicable to new users. Here, we do not aim to compare location protection methods, but to quantify the risks of realistic semantic privacy attacks without access to a TTP service. Thus, we utilize location obfuscation mainly as a tool for modelling reduced data quality in real-world scenarios. As proposed by \citet{shokri2015quantifying}, we evaluate the attacker's accuracy to quantify privacy loss.

\section*{Experimental design} 

We take a machine learning viewpoint and assume that the attacker aims to learn a mapping from visited locations to categories. The available data are a time series of location visits of a new user $u$. We group the raw data by location in order to gather temporal information about the visitation patterns to one location. The dataset $D_u$ for one user $u$ can be formalized as 
\begin{gather}
    D_u = \{\big(l_i^u, [t_1(l_i^u), t_2(l_i^u), \dots]\big)\ |\ l_i^u\in L_u\}\ 
= \{\big(l_i^u, T_u(l_i^u) \big)\ |\ l_i^u\in L_u\}\,
\end{gather}
where $L_u$ is the set of all locations visited by the user $u$, $l_i^u$ is one location in $L_u$, and $t_j(l_i^u)$ is the time of the $j$-th visit of user $u$ to location $l_i^u$. For simplicity, we abbreviate the ordered list of visit times as $T_u(l_i^u)$. Furthermore, we assume there exists an unambiguous mapping $c: L \longrightarrow C$ from each location to a category from a predefined location-category set $C$. For example, $C = \{\text{ Dining, Sports, Shopping}\}$ and the categories for user $u$ are $c(l_1^{u}) = \text{Shopping}$, $c(l_2^{u}) = \text{Dining}$, etc. 

The attacker aims to learn a model $\hat{c}$ that approximates the true mapping $c$. 
The most straightforward approach for $\hat{c}$ is a spatial nearest neighbor join with a public POI dataset; i.e., if the spatially closest POI is a restaurant, then $\hat{c}(l_i^u) = \text{Dining}$. More sophisticated methods could pool the spatial and temporal information and frame $\hat{c}$ as a machine learning model. Here, we simulate the latter via the XGBoost (XGB) algorithm~\citep{chen2016xgboost}. XGB is a tree-based boosting method that was repeatedly shown to outperform Neural Networks on tabular data~\citep{grinsztajn2022tree} and is known to perform particularly well in classification tasks with unbalanced data, as it is the case here. We also chose XGB for its interpretability and since it was empirically superior to a multi-layer perceptron approach in our tests (see Methods - Machine learning model). Together, we consider the following attack scenarios: 
\begin{itemize}
    \item \textbf{Spatial join:} For each user-location $l_i^{u}$, the category of the public POI that is closest to its geographic location $(x(l_i^{u}), y(l_i^{u}))$ is assigned. 
    \item \textbf{XGB temporal:} The attacker employs a learning approach, namely XGBoost, based on temporal information derived from $T_u(l_i^u)$ (see Methods - Temporal features). 
    \item \textbf{XGB spatial:} The attacker trains a model on spatial context features (see Methods - Spatial features). No temporal visit information is considered, only coordinates and publicly available POI data.
    \item \textbf{XGB spatiotemporal:} The model is trained on all available features, i.e., features derived from $(x(l_i^{u}), y(l_i^{u}))$ and $T_u(l_i^u)$ as well as available POI data.
\end{itemize}
In addition, we report the results for an uninformed attacker, where the predictions are drawn randomly from a categorical distribution, with the class probabilities corresponding to the class frequency in the training data.
\begin{figure*}[b!]
    \centering
    \includegraphics[width=0.95\textwidth]{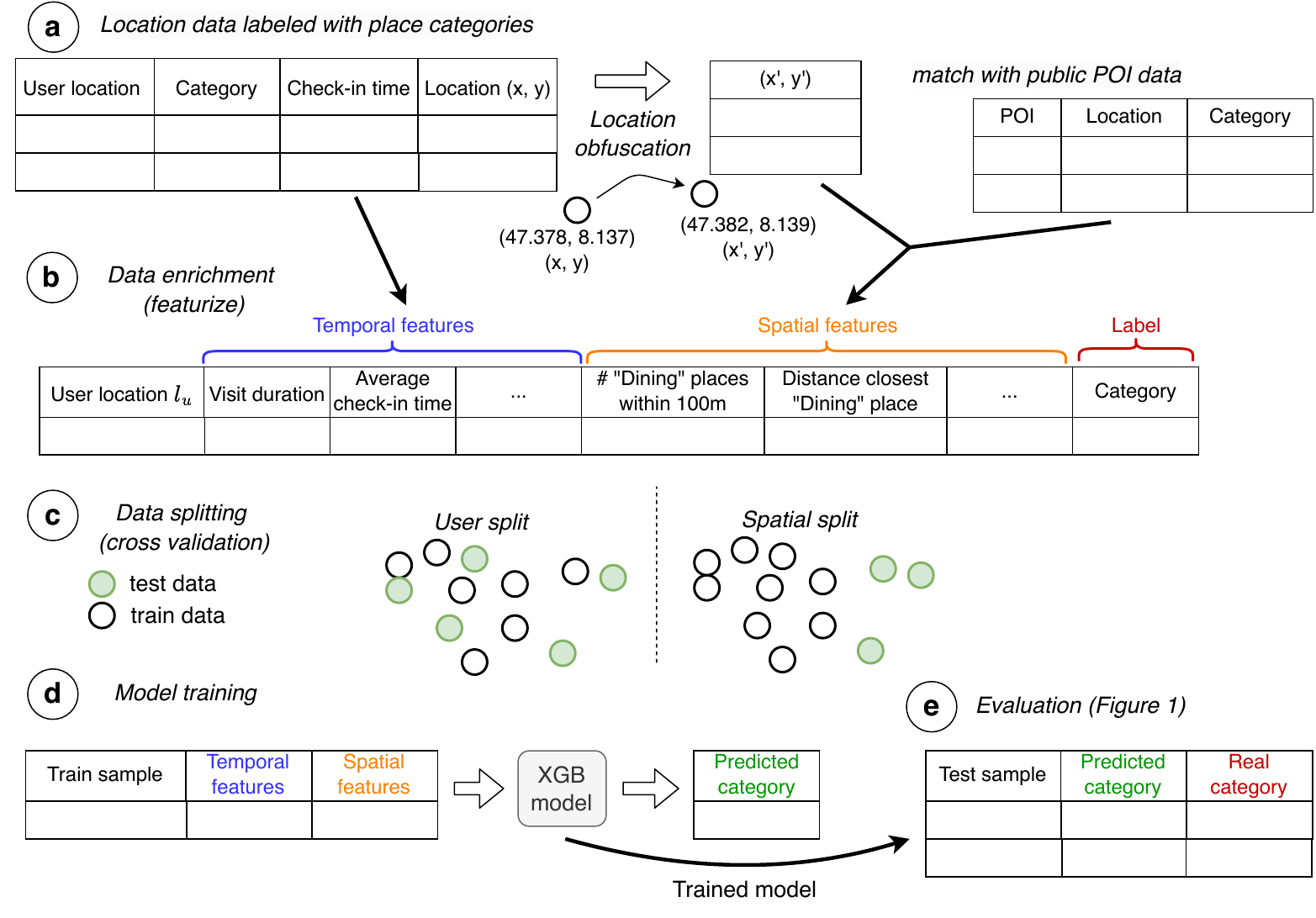}
    \caption{Overview of the experimental setup. The samples are spatiotemporal data about location visitation patterns. We simulate reduced data quality and potential protection measures by obfuscating the geographic coordinates. The samples are then featurized into vectors encoding temporal visitation patterns and spatial context. We simulate a privacy attack with a trained ML model on new users by a train-test split and evaluate the attacker's accuracy on the test data.}
    \label{fig:main}
\end{figure*}

In our experimental setup, we take an ML perspective and simulate the attack on \textit{new} users via a train-test data split. Evaluating the accuracy of this attack requires a \textit{labeled} dataset $\mathcal{D}$ of user-location pairs $l_i^u$; i.e., the location category $c(l_i^u)$ must be \textit{known}. GNSS tracking datasets usually do not provide detailed and reliably place labels. Instead, we found a public dataset from the location-based social network Foursquare most suitable for this experiments since location visits are given as check-ins to places of known categories. The dataset was already used for related tasks~\citep{ye2011exploiting, an_enabling_2022, yang2014modeling, gao2017extracting}, but without regarding privacy aspects. The places are categorized into 12 distinct classes according to the Foursquare place taxonomy (see \autoref{fig:confusion} for the list of categories and section Methods - Data and preprocessing for details). Additionally, we also use the Foursquare places as public POI data that may be exploited by the attacker as auxiliary spatial context data. 
\autoref{fig:main} provides a visual overview of the experimental setup. The input data (geographic coordinates and time points) are enriched with spatial and temporal features. Before computing spatial features, the location is \textit{obfuscated} within a varying radius $r$ to simulate GNSS inaccuracies and possible privacy protection measures (see Methods - Location masking). Then, the data is split into train and test sets, either by user or spatially, to simulate transfer to new users or even to other geographic regions. All results are reported on the combination of all test sets from 10-fold cross validation (Methods - Data split).

\section*{Results}

\subsection*{Effect of location obfuscation on place labeling accuracy}

The results for task 1 (location categorization) are evaluated in terms of accuracy, i.e., the number of correctly categorized places divided by the total number of samples, across all users and all locations (90790 samples in NYC and 211834 in Tokyo):
\begin{gather}
    Acc(\hat{c}, c) = \frac{\sum_{l_i^u \in \mathcal{D}} \mathbbm{1}[\hat{c}(l_i^u) = c(l_i^u)] }{|\mathcal{D}|} 
\end{gather}
%
\begin{figure}[b]
    \centering
    \includegraphics[width=0.6\columnwidth]{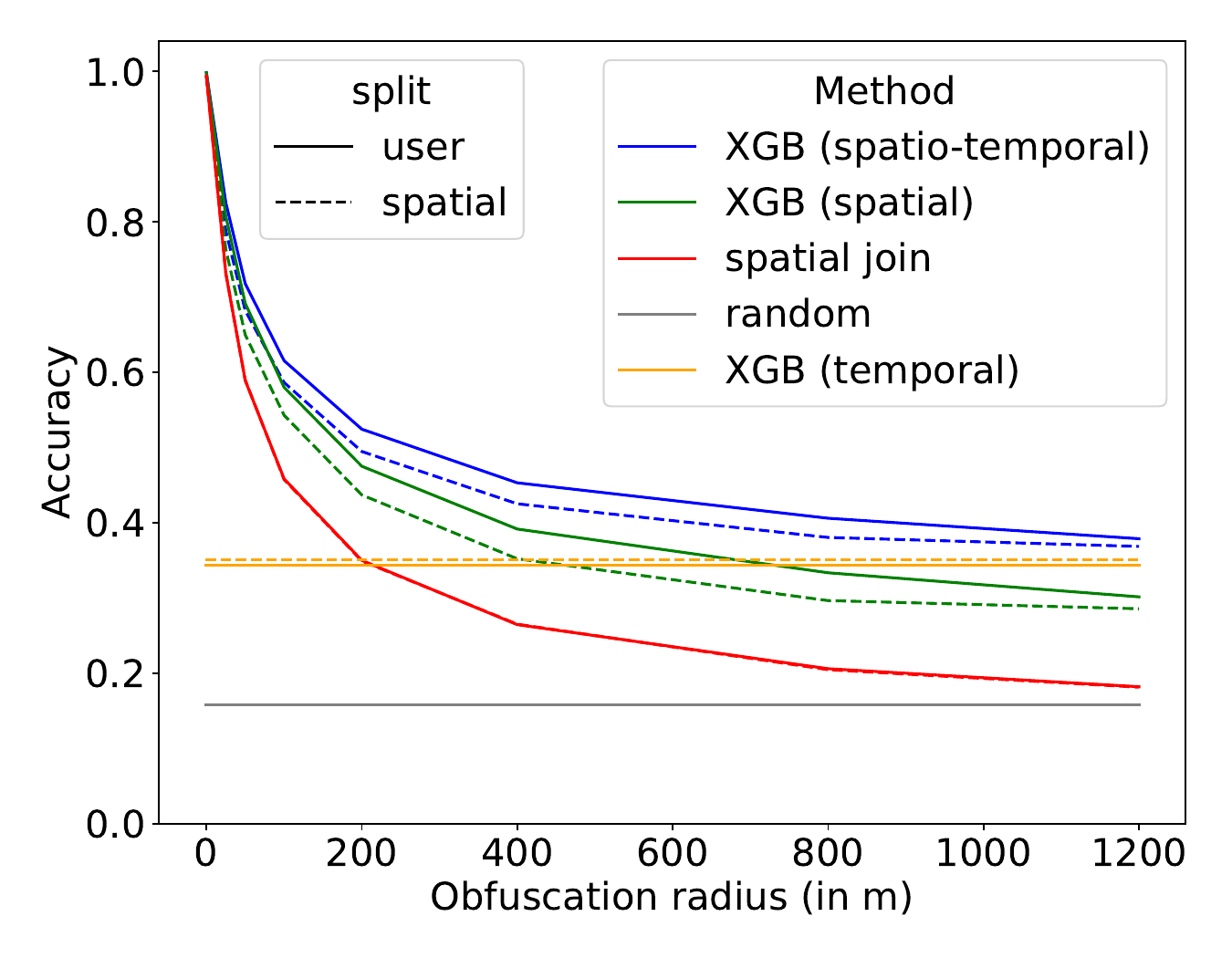}
    \caption{Effect of location obfuscation radius on the attacker's performance in categorizing locations. Spatial information are valuable for an ML algorithm even with up to 1km of obfuscation.}
    \label{fig:main_result}
\end{figure}

\begin{figure}
    \centering
    \includegraphics[width=\columnwidth]{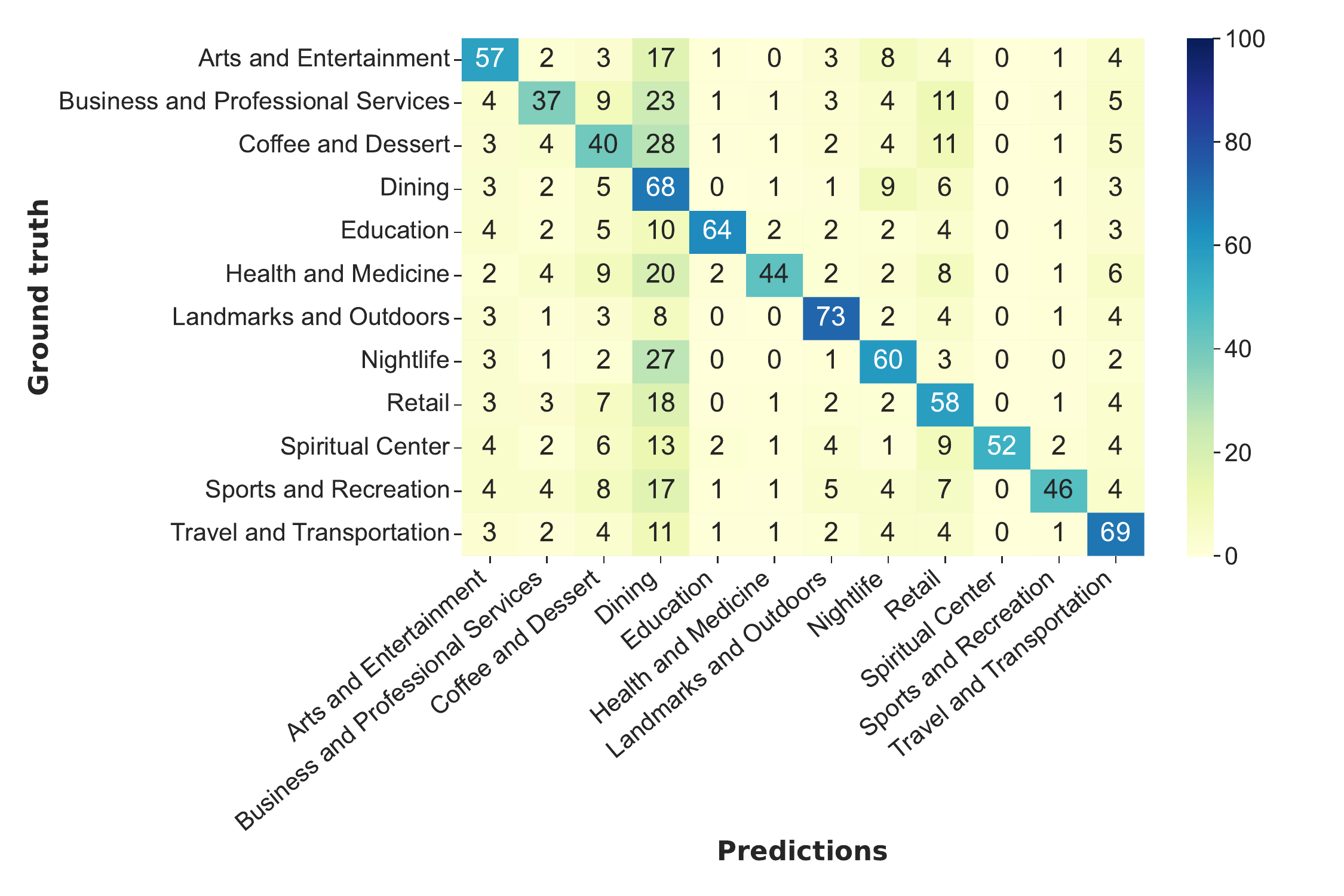}
    \caption{Normalized confusion matrix of predictions in NYC with Foursquare data and location prediction with an obfuscation radius of 100m. The accuracy is rather balanced across categories; however, many activities are erroneously classified as ''Dining''.}
    \label{fig:confusion}
\end{figure}

\autoref{fig:main_result} shows the classification accuracy of the attack scenarios by the obfuscation radius. Note that $r=0$ is an unrealistic scenario, since the check-in data and the public POI context data are both from the Foursquare dataset and are based on the exact same set of geographic coordinates. Thus, a simple spatial nearest neighbor join of the check-in location with public POIs achieves 100\% accuracy if no obfuscation is applied. Deriving a user's location from tracking data would obviously hardly yield the exact same point coordinates as a public POI. We, therefore, consider more realistic scenarios with weak obfuscation, and, additionally, protective scenarios with strongly obfuscated coordinates. The results presented in \autoref{fig:main_result} indicate that the accuracy decreases rapidly with the obfuscation radius. However, even when the attacker uses only temporal information, the accuracy is 39.1\% for Tokyo and 29.7\% for NYC, which is significantly better than random (grey line). On top of that, spatial context information can benefit the attack even when the location is obfuscated within a radius of 1km. This is remarkable and demonstrates the danger of powerful privacy attacks that make use of public POI data. In the appendix, we relate these findings to the spatial autocorrelation of place types (\autoref{fig:variogram}) and we demonstrate that the results of NYC and Tokyo are surprisingly similar (see appendix \autoref{fig:ny_tokyo}). 
Furthermore, the categorization accuracy depends on the place type; i.e., some categories are harder to detect than others. \autoref{fig:confusion} presents the confusion matrix for the attack scenario at 100m obfuscation. The error is more evenly distributed over categories than expected, although ''Dining'' and ''Retail'' are predicted disproportionally often (see appendix \autoref{fig:sensitivity_category}).

\autoref{fig:main_result} additionally compares a user split to a spatial split to analyze generalization across space (see Methods - Data split). Note that a user split is expected to be strictly better than the spatial split because the input data does not include user-identifying information such as age or gender, rendering the generalization to new users as easy as to any new samples. Surprisingly, the spatial cross-validation split only has a minor effect on the attacker's accuracy (decrease of $\sim5$\%). We conclude that the attacker's training data set is not required to cover the exact same region for the privacy attack to be successful.

\subsection*{User profiling error for probabilistic and frequency-based profiling}

\begin{figure}[ht]
    \centering
    \includegraphics[width=0.6\columnwidth]{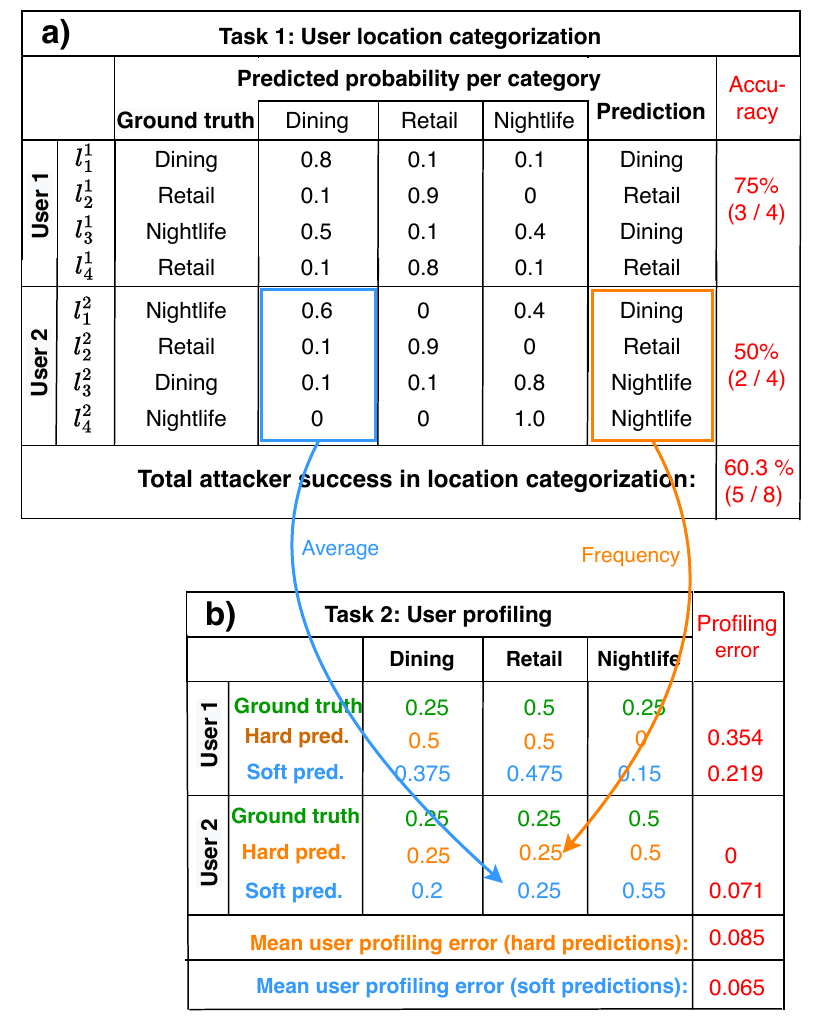}
    \caption{User profiling from location labelling. The predicted labels for individual location visits can be aggregated per user to yield an estimated user profile, either by the frequency or the average probability of the predictions.}
    \label{fig:user_profiling}
\end{figure}

While the ability of a potential attacker to categorize visited locations is concerning, we argue that the main risk is \textit{user profiling} based on the predicted categories. 
It is unclear to what extent the high categorization accuracy on a location level transfers to a high profiling accuracy on a user level. 
Here, we define a user profile as the frequency of different types of 
locations in the user's mobility patterns. Our definition 
corresponds to the term-frequency in the TF-IDF statistic\footnote{The inverse term frequency (IDF) would correspond to a weighting of the user's category-frequency by the overall frequency of this category in the data, giving higher weights to rare categories. Since the weights are the same for all users, IDF does not help to distinguish users, neither intuitively nor empirically. We therefore only characterize users by the easily interpretable TF term.}, which measures the frequency of a word in a specific document in relation to the overall occurrence of the term (in the corpus). 
Here, the ''words'' are place categories and a ''document'' is the location 
trace of one user. We provide examples for such TF-based user profiles in \autoref{fig:user_profiling}b (''Ground truth''). 
In the following, we define $p(u)$ as the profile of user $u$, and $p_c(u)$ as the entry of the vector corresponding to the frequency of category $c\in C$. For example, the ground truth profile of User 1 in \autoref{fig:user_profiling}b corresponds to $[0.25, 0.5, 0.25]$, since $p_{\text{Dining}}(\text{User 1}) = 0.25, p_{\text{Retail}}(\text{User 1}) = 0.5, p_{\text{Nightlife}}(\text{User 1}) = 0.25$. 
In this study, we aim to quantify how accurately the adversary could predict $p(u)$. The evaluation of user profiling performance boils down to comparing the difference between two categorical distributions, namely the distributions of the real profile $p(u)$ versus the predicted category frequencies $\hat{p}(u)$: 
\begin{gather}
    E_{\hat{p}(u), p(u)} = \sqrt{\sum_{c\in C} (\hat{p}_c(u) - p_c(u))^2}
\end{gather}

\begin{figure}[ht]
    \centering
    \includegraphics[width=0.6\columnwidth]{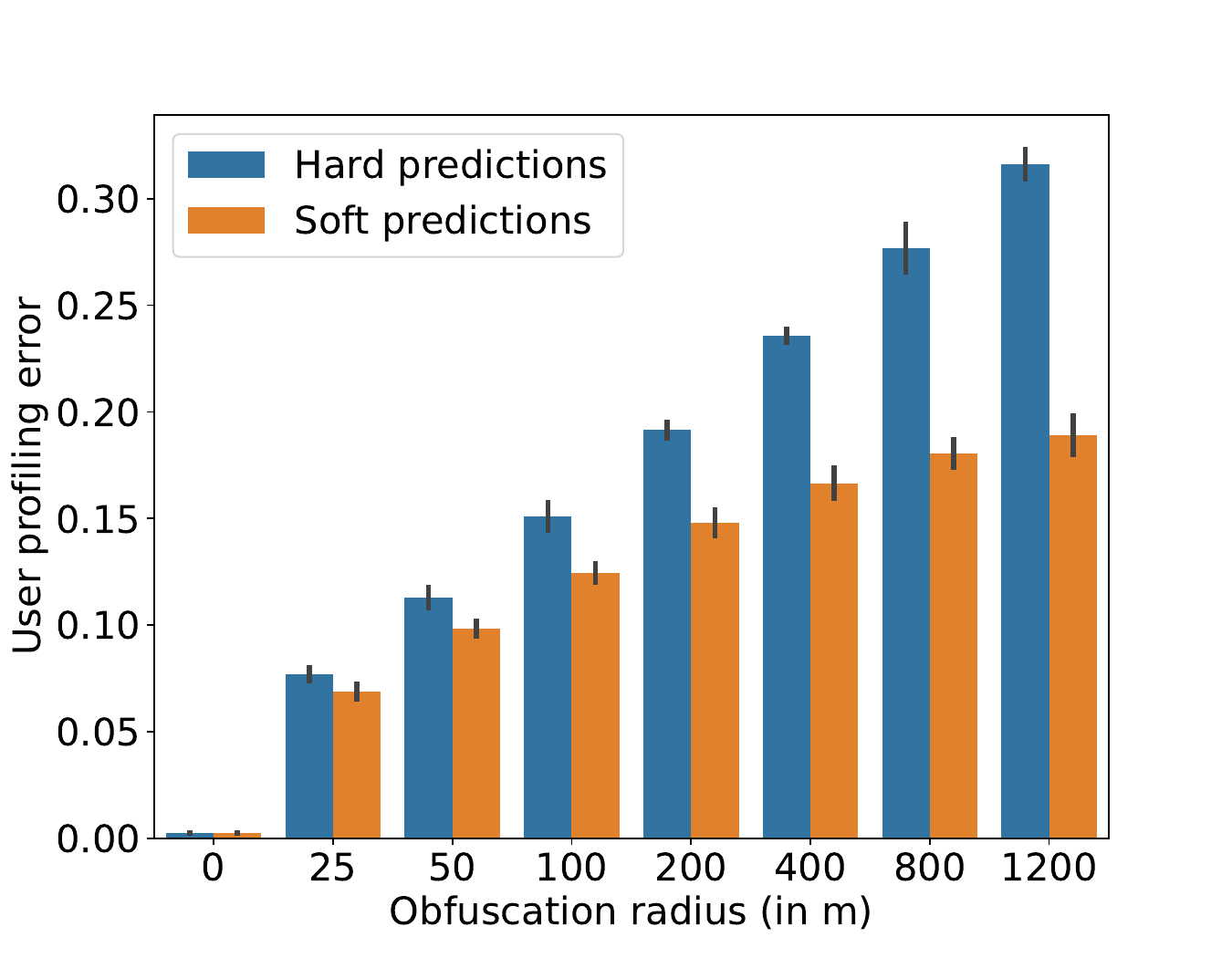}
    \caption{Comparison of user-profiling errors achieved from averaging ''hard'' predictions or ''soft'' prediction probabilities for each category. Probabilistic classifications improve the spatial attack, in particular for lower-quality location data.}
    \label{fig:hard_soft_labels}
\end{figure}

The attacker can estimate the profile $\hat{p}(u)$ simply by counting the predicted place categories. For example, in \autoref{fig:user_profiling} the ''Retail'' category is predicted one out of four times for user 2 and therefore takes a value of 0.25 in the profile (see orange arrow). However, many ML-based classification models actually predict a ''probability''\footnote{The probability distribution over categories is usually derived from the predicted values with a softmax function or by averaging hard predictions of base estimators and is, therefore, by no means the actual posterior distribution. While the provided uncertainties are oftentimes poorly calibrated~\citep{guo2017calibration}, they nevertheless add information to the final predicted label.} for each category, as shown in \autoref{fig:user_profiling}a. The XGBoost model, for example, outputs the prediction frequency of each category among its base learners (decision trees). Probabilistic predictions provoke a second way to estimate $\hat{p}(u)$, namely by averaging the predicted probabilities per category (see blue arrow in \autoref{fig:user_profiling}).
In the following, we term the first option (computing the frequency of predicted categories, orange) as ''hard'' profiling and the second option (averaging category-wise probabilities, blue) as ''soft'' user profiling. As shown in the toy example in \autoref{fig:user_profiling}, soft profiling can increase or decrease the error compared to hard profiling (e.g., decrease from 0.354 to 0.219 for user 1, but increase from 0 to 0.071 for user 2).

In \autoref{fig:hard_soft_labels}, we empirically compare both strategies on our dataset in terms of the error $E$ defined above. 
Only the error for the strongest attack scenario (XGB spatio-temporal) is shown, averaged over cities (NYC and Tokyo). 
The profiling error is significantly lower for the soft profiling strategy that is based on probabilistic predictions. In particular, the error of ''hard'' profiling increases proportionally with a doubling of the obfuscation radius, while the error of soft-labeling increases sub-linearly (see \autoref{fig:hard_soft_labels}). This result is consistent for all considered scenarios. It demonstrates that well-calibrated probabilistic prediction methods are more dangerous in terms of user profiling than point predictors, even if the latter may achieve a higher place classification accuracy. 

All further results are reported for the \textit{soft} predictions in order to simulate the strongest attack.

\subsection*{User reidentification accuracy based on the estimated profiles}

Judging from the error alone it is difficult to interpret how much the user profile actually reveals. Such interpretation depends on the variance of the user profiles: For example, if all users have the same profile, the prediction error may be very low, but there is no value in profiling. As a more interpretable metric, we follow previous privacy research and analyze the possibility of re-identifying users by their predicted profile. Given the pool of ground-truth user profiles (\autoref{fig:user_profiling}b green), we match the predicted profiles by finding their nearest neighbors in the pool based on the Euclidean distance of their profile vectors. We report the results in terms of top-5 re-identification accuracy, also called hit@5. 

\begin{figure}[hbt]
\begin{minipage}{0.48\textwidth}
    \centering
    \includegraphics[width=\columnwidth]{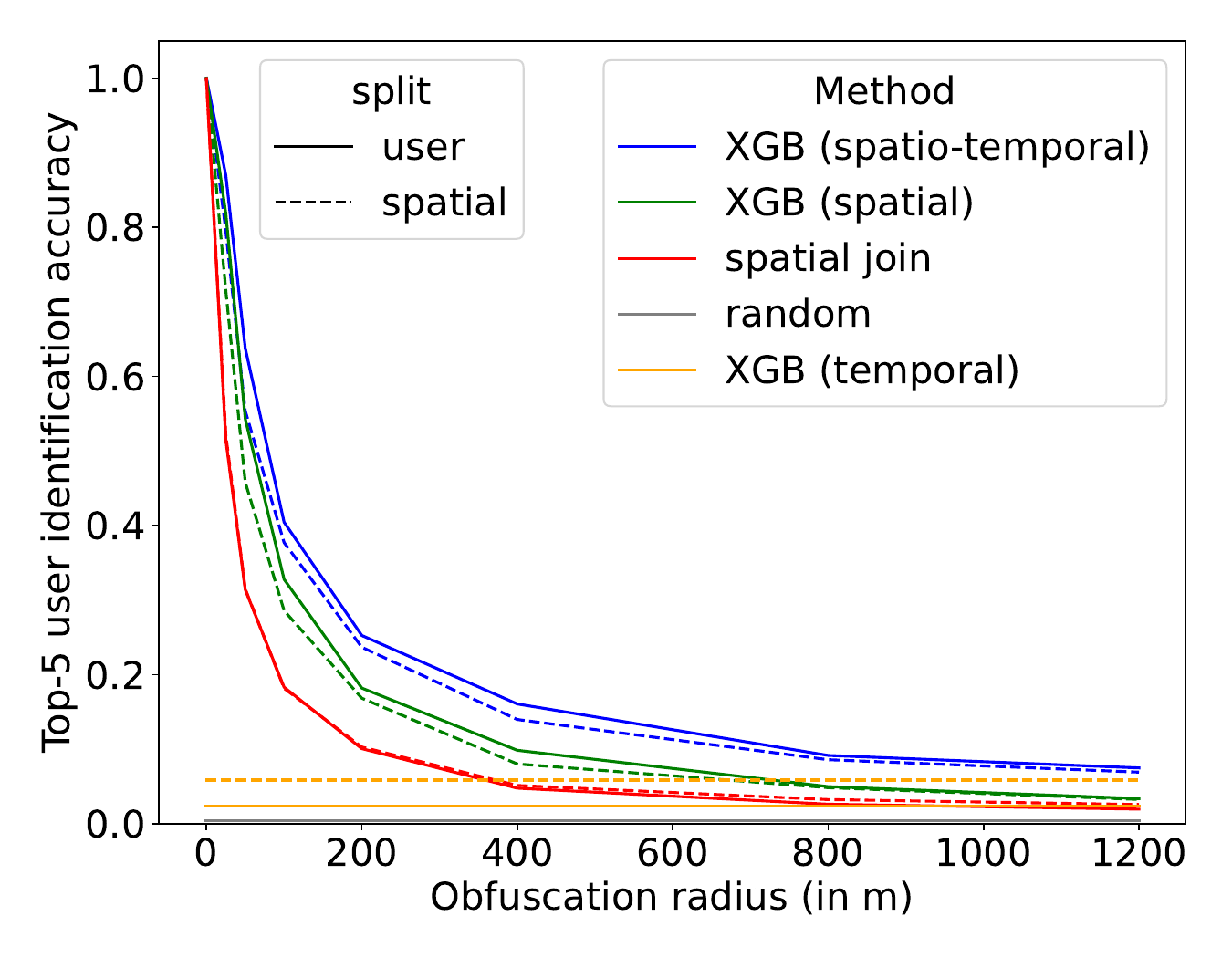}
    \caption{User-profiling performance of different semantic attacks, in terms of the top-5 accuracy of re-identifying users by their profile. With an obfuscation radius of around 400m, the user profiling accuracy converges.}
    \label{fig:user_profiling_results}
\end{minipage}
\hfill
\begin{minipage}{0.48\textwidth}
    \centering
    \includegraphics[width=\columnwidth]{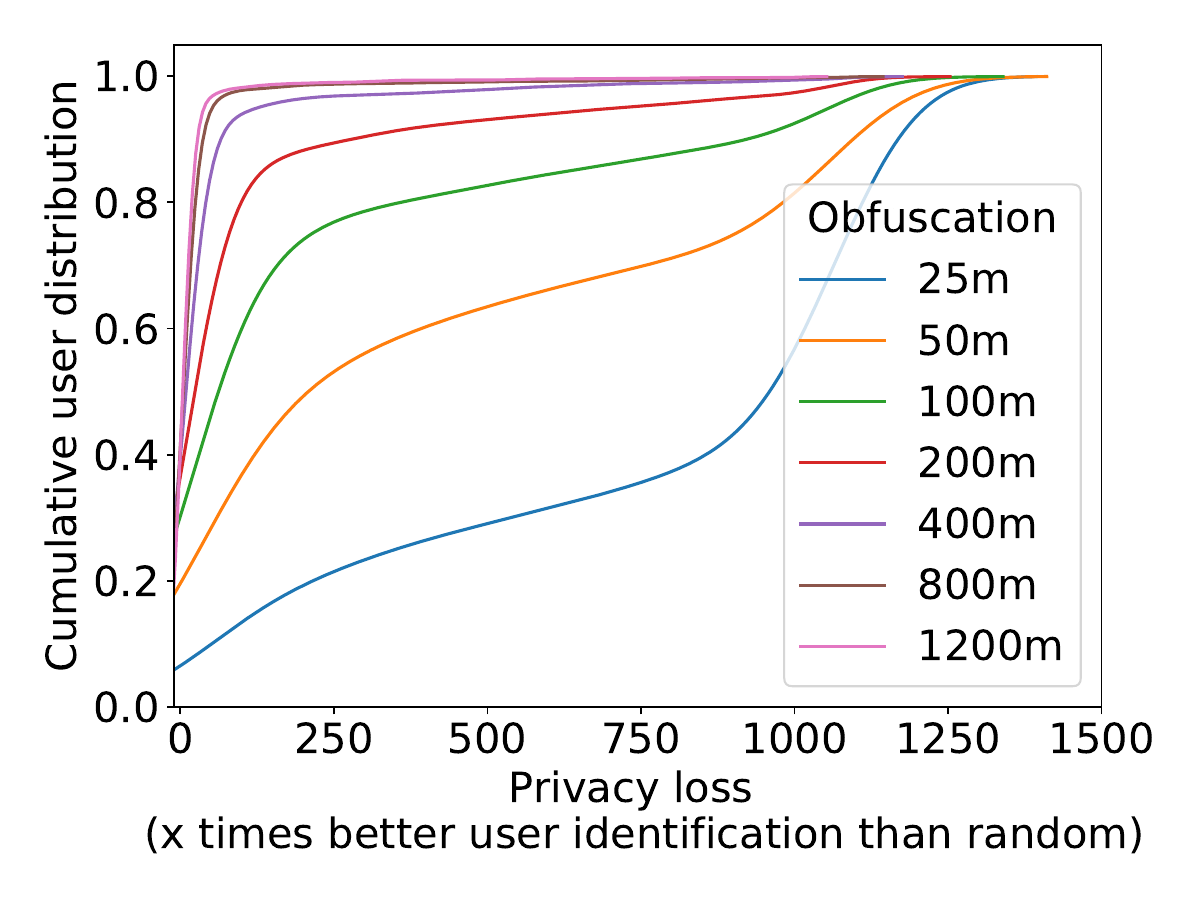}
    \caption{Cumulative distribution of the privacy loss per user caused by the strongest attack scenario.}
    \label{fig:priv_loss_dist}
\end{minipage}
\end{figure}

In \autoref{fig:user_profiling_results}, the re-identification accuracy is shown by the attack scenario. A corresponding plot of the profiling error is given in the appendix (\autoref{fig:user_mae}). Although the accuracy decreases quickly with stronger obfuscation, it is still larger than 10\% even with an obfuscation radius of 1.2km. The average uninformed (random) identification accuracy is 0.6\% on average, with 1083 users in NYC and 2293 users in Tokyo. To compare the decay of the user profiling performance to the decay in place categorization accuracy (\autoref{fig:main_result}), we fit an exponential function of the form $f(x) = a + c \cdot e^{-x\cdot \lambda}$ to both results. The place categorization accuracy decays with $a=0.3439, \beta=0.0097, c= 0.6216$, indicating that the accuracy decreases with a rate of $e^{-0.0097} = 0.9903$ but converges to around $0.3439$. The function fit for the user identification accuracy yields $a=0.0625, \beta=0.0121, c= 0.9518$. In other words, with every 50 meters added to the location obfuscation radius, the user re-identification accuracy is reduced by a factor of $0.5488$ ($=e^{-0.0121 * 50}$). With $r = 57.43$, the accuracy has approximately halved. This firstly demonstrates that place categorization does not directly translate into user profiling, as the profiling accuracy decays faster than the categorization accuracy, and secondly gives guidance for selecting a suitable masking radius.  

\subsection*{Induced privacy loss of ML-based privacy attacks}

Finally, we transform the re-identification accuracy into a \textit{privacy loss} metric following \cite{manousakas_quantifying_2018}. They define the privacy loss $PL$ for one user $u\in U$ as 
\begin{gather}
PL(u) = \frac{P_{attack}\big(u = u^*\ |\ D_u\big)}{P_{uniformed}(u = u^*)}
\end{gather}
where $P_{uniformed}$ is the probability of an uninformed adversary to match $u$ to the true user $u^*$, 
corresponding to a random pick from all users $U$, so $P_{uninformed} = \frac{1}{|U|}$. 
The probability of an informed adversary, on the other hand,  is the probability to match the user to the correct profile by utilizing sensitive user data including geographic coordinates and visitation times. We assume that given a pool of users $U$, the attacker would match $u$ to a user $u_i\in U$ from the pool with a probability proportional to the similarity of their profiles: 

\begin{gather}
P_{attack}(u = u_i | \mathcal{D}) \propto softmax\big(sim(u, u_i)\big) =
\frac{e^{sim(u, u_i)}}{\sum_{j=1}^{|U|} e^{sim(u, u_j)}}
\end{gather}
where we define the similarity as the inverse distance of the user profile vectors $sim(u, u_i) = \big(E_{\hat{p}(u), p(u_i)} \big)^{-1}$. Note that \citet{manousakas_quantifying_2018} use a rank-based measure of similarity, which however seems unintuitive given that we know the exact distance between each pair of user-profiles and not only their respective rank. 

The median privacy loss is $11$ if the adversary is given spatio-temporal information where the locations are obfuscated by 100m (see appendix \autoref{tab:results_100}). In other words, the adversary is still 11 times better at re-identifying a user by his profile than with a random strategy. Moreover, the adversary with spatio-temporal data is $9.9$ times better than an adversary that uses only temporal information, even though the spatial data are obfuscated up to 100m. At higher location obfuscation, the privacy loss converges. The strongest attack only yields a median privacy loss of $3.74$ at 200 meters obfuscation radius and $2.13$ at 400m. However, the privacy loss strongly varies across users. 
\autoref{fig:priv_loss_dist} shows the cumulative distribution of users. If the locations are obfuscated by 100m, around 80\% of the users have a privacy loss lower than 250; however, the distribution is heavy-tailed with a considerable number of users that are still easy to identify. 
Nevertheless, we conclude that obfuscating the location with a radius between 100 and 200 meters would significantly reduce the risk of successful profiling attacks for a large majority of users.


\FloatBarrier 
\subsection*{Features that affect the predictability of place categories}

\begin{figure}[htb]
    \centering
    \begin{subfigure}[b]{0.975\textwidth}
    \includegraphics[width=\columnwidth]{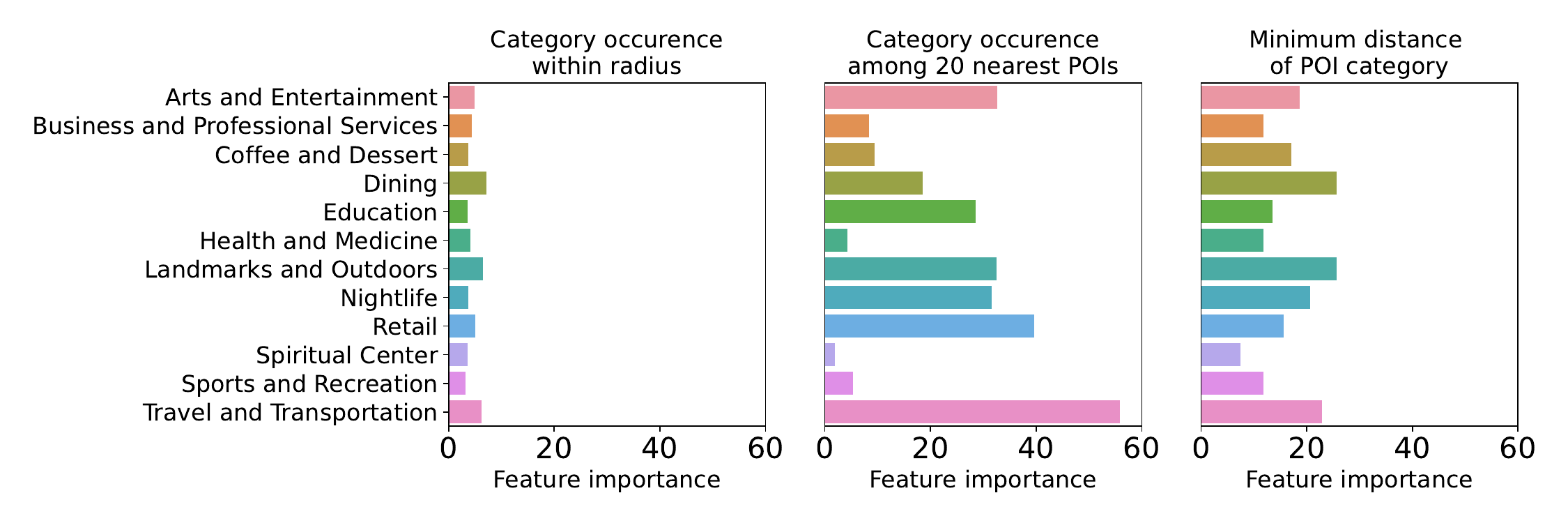}
    \end{subfigure}
    \begin{subfigure}[b]{0.975\textwidth}
    \includegraphics[width=\columnwidth]{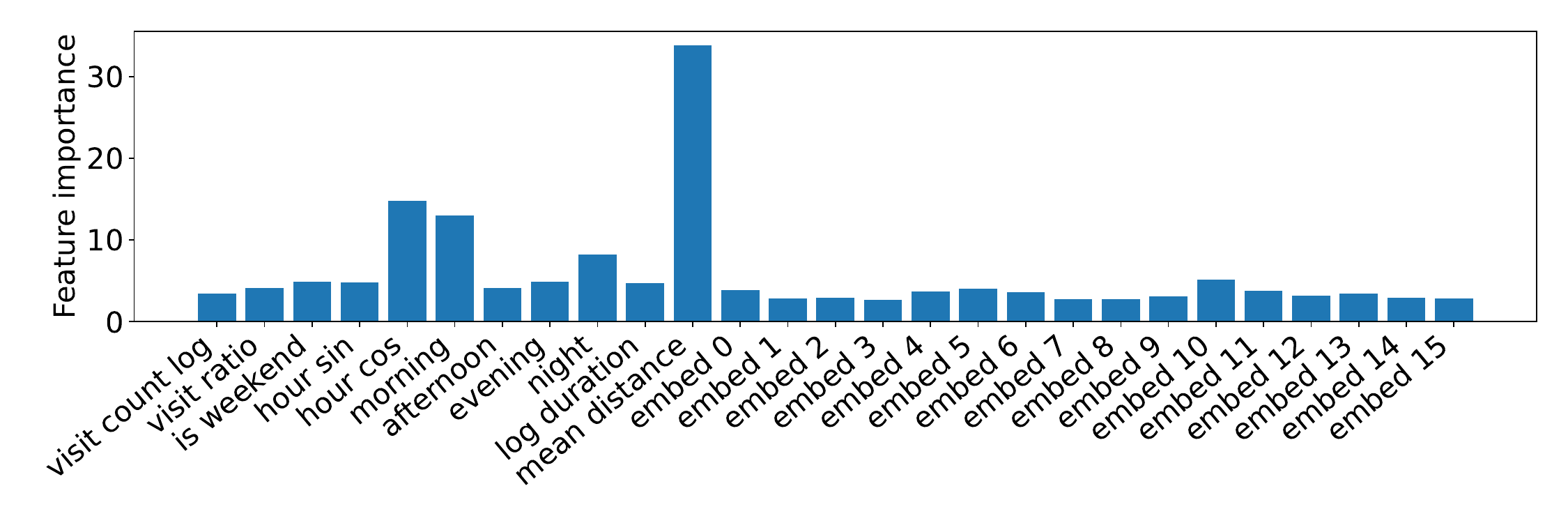}
    \end{subfigure}
    \caption{Feature importances in the XGBoost classifier. The occurence of different categories and their mean distance are the most important features for place categorization.}
    \label{fig:importance}
\end{figure}

One advantage of boosted-tree based machine learning methods such as XGBoost is that decision trees are interpretable. While the individual decision boundaries are not transparent in large ensembles of trees, one can still compute the importance of individual features in terms of their mean decrease of data impurity. The respective importance of the spatial and temporal features included in our study are shown in \autoref{fig:importance}. The most important spatial features are the number of POIs per category among the $k$ nearest POIs. The spatial embedding features derived with the space2vec (embed 0 - embed 16) method apparently do not add much information. The time of the day, expressed in sinus and cosinus of the hour and binary variables for morning, afternoon and evening, also play a significant role, highlighting the relevance of temporal information.

\FloatBarrier

\subsection*{Dependency on POI data quality}

\begin{figure}[htb]
\begin{minipage}{.43\textwidth}
    \centering
    \includegraphics[width=\columnwidth]{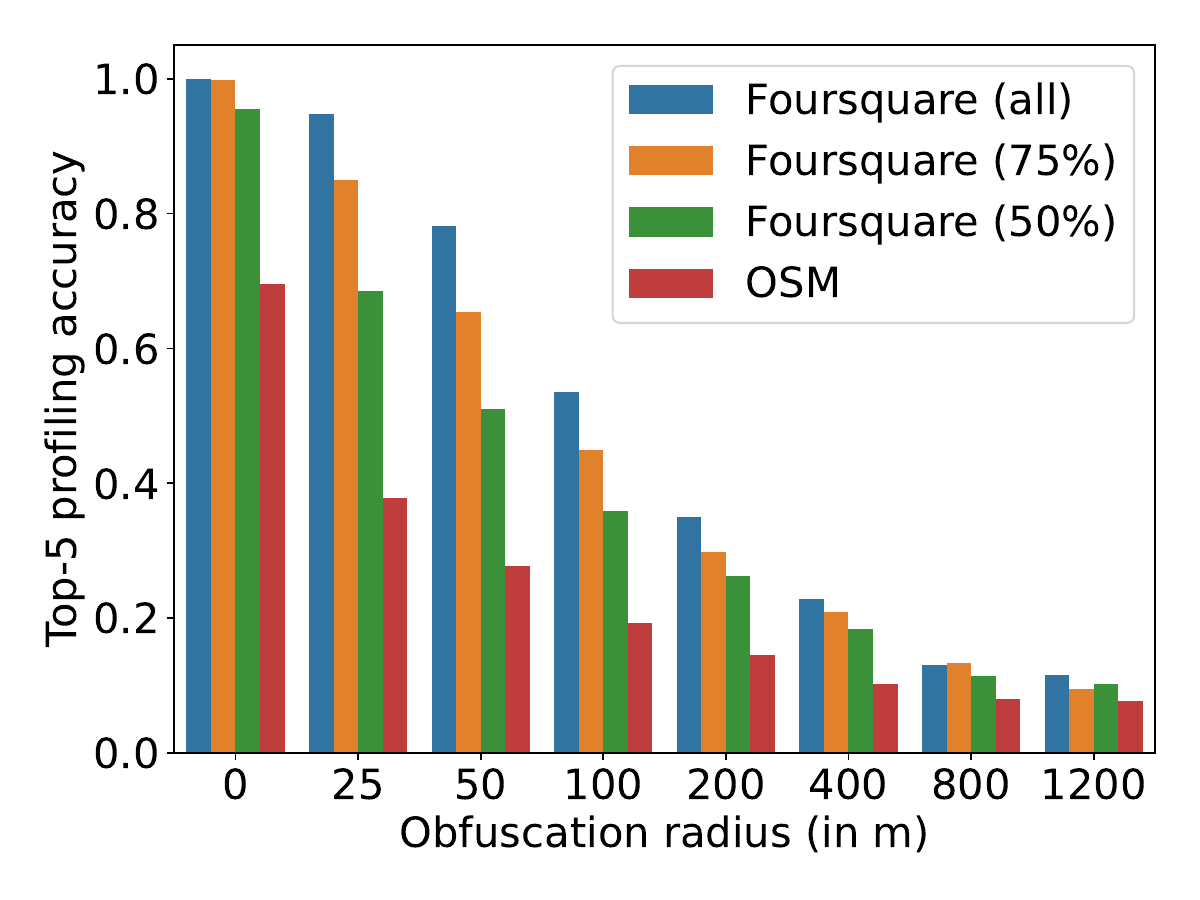} 
    \caption{Dependency of the attacker's success on the POI quality. The strongest attack scenario based on spatio-temporal data is shown. While the completeness of POI data has a disproportionally low impact on user profiling, using OSM data decreases the attacker's success.}
    \label{fig:poi_quality}
\end{minipage}
\hfill
\begin{minipage}{.53\textwidth}
    \centering
    \includegraphics[width=\columnwidth]{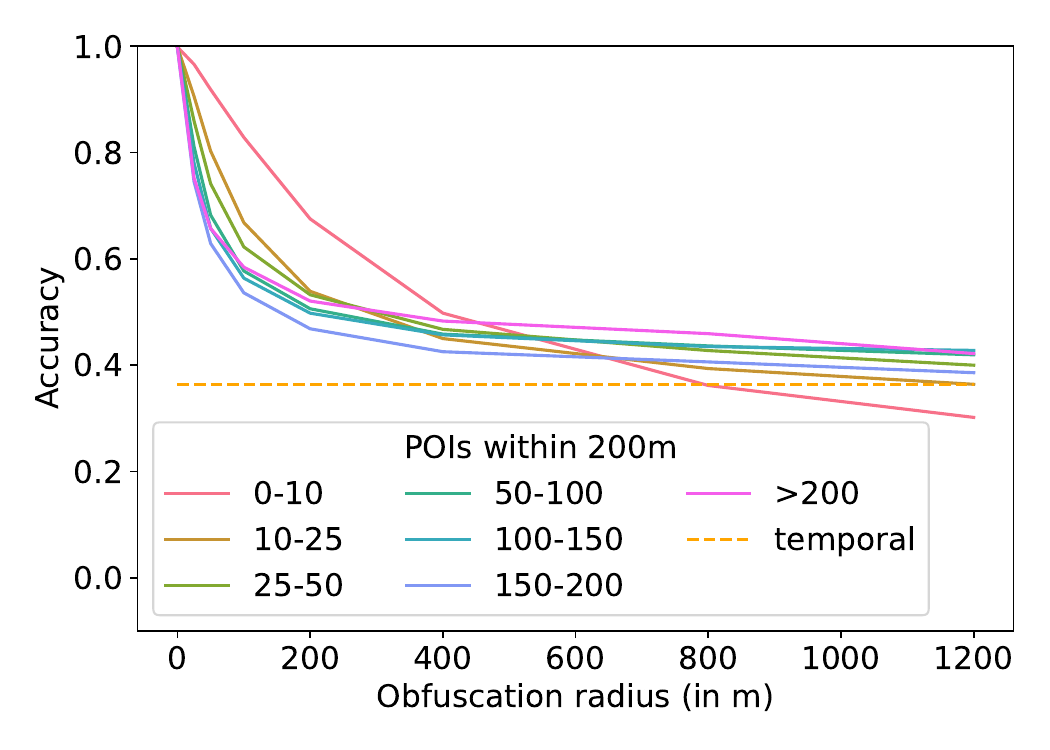}
    \caption{Place categorization accuracy by POI density (number of POIs within 500m). Visited places in very dense areas are harder to classify.}
    \label{fig:density}
\end{minipage}
\end{figure}

To simulate incomplete POI data, we subsample 75\% or 50\% randomly from the Foursquare POIs. Furthermore, the performance with POI data from OSM instead of Foursquare is evaluated. In this experiment, only the predictions of the strongest attack (XGB spatio-temporal) on NYC check-in data are evaluated. \autoref{fig:poi_quality} depicts the results, where ''Foursquare (all)'' corresponds to the results in \autoref{fig:main_result}. The removal of Foursquare POIs has surprisingly little effect on the user identification accuracy. Even with 50\% of the POIs, 84.8\% of the check-ins can be classified correctly (see appendix~\autoref{fig:poi_user}), translating to a top-5 identification accuracy of 94\%. This is due to the spatial autocorrelation between places of certain categories (see appendix~\autoref{fig:variogram}).

Meanwhile, it is much harder to classify the category of Foursquare check-ins with OSM POIs. We hypothesize that this is due to substantial differences between OSM and Foursquare POI data. Previous work \cite{zhang2019using} tried to match cafes in the OSM dataset to cafes in the Foursquare set and find that only around 35\% can be matched exactly (Levenshtein distance of labels=1), with a spatial accuracy of around 30-40m. 
In addition to these location differences, in our case there are also differences in the place categories, which we partly had to assign \textit{manually} to the OSM POIs (see Methods - Data and preprocessing). Nevertheless, the low performance with OSM data unveils important difficulties for an attacker to utilize inaccurate, incomplete and dissenting datasets of POIs.

\subsection*{Influence of the POI density}

Furthermore, the difficulty level of the attack depends on the density of spatial context data, since it is easier to match a location to a nearby POI if the number of nearby POIs is low. We quantify this relation by computing the number of surrounding POIs within 200m for all considered places in NYC and Tokyo. In \autoref{fig:density}, the place labelling accuracy is shown by POI density groups. Places in dense areas; i.e., with many surrounding POIs, are harder to classify. For example, when the obfuscation radius is 100m, the mean number of POIs within 200m around the (non-obfuscated) location is 58 for correctly predicted samples, but 85 for erroneously classified samples. However, the variance between the curves shown in \autoref{fig:density} is lower than expected. Only points with less than ten nearby POIs are significantly easier to match.


The dependence of the predictability on the POI density calls for a context-aware protection scheme~\citep{zhang2018context, andres2013geo}. We implement such scheme by setting the obfuscation radius $r$ for a specific location such that at least $m$ public POIs lie within the radius. For the sake of comparability, we tune $m$ to a value that leads to an average obfuscation radius of $200m$ ($m=16$). In other words, when obfuscating each location $l$ within a context-aware radius $r(l)$ that covers exactly 16 public POIs, then $\frac{1}{|\mathcal{D}|} \sum_{l\in \mathcal{D}} r(l) \approx 200$. As desired, this masking scheme destroys the relation between POI density and accuracy. However, our experiments show that the average accuracy \textit{increases} compared to the accuracy reported for location-independent masking in \autoref{fig:main_result} (accuracy of $0.52$ compared to $0.49$ for the experiment on NYC-Foursquare data with XGB spatio-temporal). This also holds at a user-level, where the user-profiling performance is higher with context-aware location obfuscation ($0.27$ vs. $0.23$). It seems that the weak obfuscation of locations in high-density regions has a greater effect than the strong obfuscation of isolated places. We conclude that simple context-aware obfuscation based on POI density is not sufficient to reduce privacy risks, at least not at the same average obfuscation level. While the evaluation of protection methods is out of the scope of this work, further work is needed to understand their effectiveness against undesired user-profiling. 

\section*{Discussion}

We have quantified the risks of undesired user profiling in different attack scenarios, varying 1) the information available to the attacker, 2) the location data quality in terms of obfuscation radius, and 3) the POI data quality. We comment on each aspect in the following.

First, our experiments reveal that machine learning methods can efficiently exploit spatial context data, even with low data quality or incomplete data. We further confirm previous findings by \cite{mckenzie2015also} that even only temporal information about location visits poses a significant privacy risk. This risk may be further increased, for example, if also the opening times of surrounding POIs are used as input features~\citep{yan2022perturb}. In general, more powerful ML methods may increase privacy risks beyond our results. 
A particularly interesting finding 
is the superiority of \textit{probabilistic} predictions for deriving user profiles. In other words, a potential attacker can estimate the importance of different place types in a user's life without knowing the category for each individual place exactly.

Furthermore, we took a user-centric viewpoint and derived location protection recommendations. 
The exponential decay of user identification accuracy demonstrates the high effectiveness of simple protective measures, and the results suggest that the privacy risks become negligible when the location is obfuscated with a radius of around $200$m. While such inaccuracy may be intolerable in navigation apps, it yields a good trade-off in other applications such as social media, where the approximate location is still interesting to friends but not yet informative for profiling attacks. 
However, further experiments on other datasets are necessary to validate the results. Our analysis is based on an experimental setting where each visited location can for sure be matched to a public POI. An attack that aims to classify user activities that are not related to public POIs is, therefore, expected to be more difficult (e.g., detecting a visit to a friend's place). 
In the appendix (\autoref{fig:yumuv}), we provide a study on a GNSS-based tracking dataset where stay points are labeled with a few broad activity categories, but it would be highly interesting to reproduce our results on a GNSS dataset with more detailed place categories. However, datasets that are large and labeled at the same time are rare~\cite{chen2016promises}. 
%
%
%
Finally, we see a strong dependency of the attacker's success on the density and completeness of spatial context data. Thus, future privacy protection algorithms should not only regard past studies on protection efficiency, but also improvements in public databases. %
We hope to inspire future research on the risks and, importantly, on suitable protection methods against such novel semantic privacy attacks. Further analysis may, for example, investigate which users are particularly easy or hard to profile. The classification of users into a predefined set of profiles or a cluster of profiles could provide further insights into the actual dangers of unwanted behavior analysis. Finally, it may be an interesting endeavor to develop location protection techniques that specifically target the weaknesses of machine learning models, similar to adversarial attacks~\citep{huang2017adversarial}.

\section*{Conclusion}

Semantic privacy deserves more attention in geoprivacy research, considering the business case of data brokers and the interest of companies in semantic information in contrast to raw data. Our analysis is a first step towards a better understanding of the actual risk for a user to reveal sensitive behavioral data when sharing location data with applications. Spatial and temporal patterns in location data lead to a significant opportunity for user profiling, even if the coordinates are not accurate. However, this effect diminishes with stronger location protection. Our analysis, therefore, enables users and policy-makers to derive recommendations on a suitable protection strength. %

\section*{Methods}

In the following, our methods are described in detail. Our implementation is available open-source at \url{https://github.com/mie-lab/trip_purpose_privacy}.

\subsection*{Data and preprocessing}

\subsubsection*{Check-in data from Foursquare}

Our study mainly uses data from the location-based social network \textit{Foursquare}. In contrast to tracking datasets or data from other social networks (e.g., tweets), the Foursquare dataset offers labeled and geo-located place visitation data. Specifically, users check-in at venues, e.g., a restaurant, and the geographic location of the venue as well as a detailed semantic label, e.g., ''Mexican restaurant'', are known. Similar to other studies~\citep{yang2014modeling, yang2016privcheck}, we use the Foursquare subset of New York City and Tokyo in order to simplify location processing and to study the variability of the results over two different cities. The data was collected by~\cite{yang2014modeling} from 12 April 2012 to 16 February 2013 and was downloaded from their website\footnote{\url{https://sites.google.com/site/yangdingqi/home/foursquare-dataset?pli=1}}. Note that Foursquare has changed over the years, and the data thus differs from today's usage of this LBSN. This is not an issue for our study, as the underlying location visitation patterns are expected to remain similar.

As a first step, we clean the category labels of place check-ins of users. We focus on leisure activities and do not consider home and work check-ins for several reasons: 1) Home and work location can be inferred by \textit{temporal} features such as the time of the day and visit duration. Spatial POI data are not necessary. 2) Identifying home and work is possible with simple heuristics, e.g., assigning the most often visited location as home and the second-most-frequently visited location as work. We believe that previous attempts on this task mainly suffer from insufficient data quality and the lack of reference data, and not the difficulty of the task itself. 3) Many Foursquare users in the dataset do not check-in at home or work since the social network was mainly used to share leisure activities, at least in 2012 when the data was gathered and before changes where made to their (check-in) app.

%
In total, the Foursquare POIs in NYC and Tokyo are labeled with 1146 distinct categories. A taxonomy 
is provided with 11 groups on the highest level, such as \textit{Dining and Drinking} or \textit{Arts and Entertainment}. We use this categorization as the ground-truth location categories, but make a few changes in order to sufficiently distinguish common types of leisure activities that are relevant for user profiling. Specifically, we divide the category \textit{Dining and Drinking} into categories \textit{Dining} (all kinds of restaurants), \textit{Nightlife} (bars), and \textit{Coffee and Dessert}, based on the label given on lower levels of the taxonomy. Furthermore, the category \textit{Community and Government} is split into the categories \textit{Education} and \textit{Spiritual Centers}. Other subcategories that can not be fitted into these two, e.g. ''government building'' or ''veteran club'', are omitted. 
Finally, there are around 100 labels in the NYC-Tokyo Foursquare dataset from 2012 that do not appear in the (up-to-date) Foursquare POI taxonomy. We manually assign these labels to categories. The final distribution of the labels in NYC check-ins is shown in the appendix in \autoref{fig:checkins_foursquare}.

Furthermore, the check-in dataset is cleaned by merging subsequent check-ins of the same user at the same location. A check-in event is deleted if it occurs within one hour of the previous check-in at that location, leading to the removal of 0.496\% of the NYC check-ins and 0.63\% of the ones in Tokyo.

\subsubsection*{Public POI data}

We assume that the attacker can access public POI data, such as the POIs from Foursquare. However, categorizing check-in locations in the Foursquare data is easy when the Foursquare POIs are given since they correspond exactly in their geographic location and each check-in can (in theory) be matched to a known POI. Apart from obfuscating the check-in location to simulate inaccurate GNSS data, we also simulate incomplete POI data by sampling 50\% and 75\% of the Foursquare POIs at random.

%

Last, we simulate a situation with substantially different POI data by using POIs from OSM. 
The Python package \texttt{pyrosm}~\citep{pyrosm} is used to download all places of the categories ''healthcare'', ''shop'', ''amenity'', ''museum'', ''religious'', ''transportation'', and ''station'' (public transport) from OSM. The ''amenity'' category in particular contains a large collection of places, and we first delete all places labeled as ''parking space'' since they accounted for a large fraction of the data and are irrelevant to our analysis. We further manually re-label the POIs in order to assign place categories. The same categories as in the Foursquare dataset are used and the mapping from OSM-POI-types to our categories is given in detail in our code base\footnote{\url{https://github.com/mie-lab/trip_purpose_privacy/blob/main/data/osm_poi_mapping.json}}.

\subsection*{Spatial and temporal input features to machine learning model}

\subsubsection*{Temporal features}\label{sec:temporal}

Temporal features are computed from $T_u(l_i^u)$ as the following:
\begin{itemize}
   \item \textbf{Visit frequency features:} The absolute visit frequencies of location $l^u_i$, corresponding to $|T_u(l_i^u)|$, and the relative frequency with respect to all check-ins by $u$, formally 
   \begin{gather}
       \text{f}_{\text{visit\_frequency}}(l_i^u) = \frac{|T_u(l_i^u)|}{\sum_{l^u_i \in L_u} \sum_{t_j \in T_u(l_i^u) } t_j}
   \end{gather} 
   The absolute frequencies are scaled with a logarithm to reflect well-known power-law properties of location visitation patterns~\citep{brockmann_scaling_2006, rhee_levy-walk_2011}.
   \item \textbf{Duration features:} In the Foursquare dataset used as training data, the check-outs of location visits are not provided, so only the start time is known. Thus, we approximate the visit duration by computing the time until the next check-in. Since no check-outs are (publicly) available, there are many outliers with gaps over more than a day. We flatten these outliers by scaling logarithmically, and finally, we take the average over the individual visit durations. Formally, the visit time is subtracted from the time of its subsequent check-in, given as the minimum time of all following check-ins of the user:
   \begin{gather}
       \text{f}_{\text{dur}}(l_i^u) = \frac{1}{|T_u(l_i^u)|} 
       \sum_{j=0}^{|T_u(l_i^u)|} \log{\Big(\min_{\substack{k, m \\ \text{s.t. } t_m(l_k^u) > t_j(l_i^u)}}t_m(l_k^u) - t_j(l_i^u)\Big)}
   \end{gather}
   The duration of the last check-in overall is omitted. 
   Although this approximation is very rough due to the dependence on the LBSN usage frequency of users, we empirically observed that it is still helpful for inference.
   \item \textbf{Daytime features:} Last, the start time is represented by a variety of features: Binary features to indicate whether it is on the weekend, in the morning (before 12pm), in the afternoon (12pm - 5pm), in the evening (5pm - 10pm), or at night (10pm - midnight). The time thresholds were selected to reflect different activities (e.g. dining vs nightlife). The exact daytime was encoded with trigonometric functions (sine and cosine) to reflect their cyclical properties, as is common in machine learning.
\end{itemize}

\subsubsection*{Spatial features}\label{sec:spatial}

The attacker can utilize the recorded geographic coordinates to predict the location category. However, inputting the raw coordinates to a model is not advisable as they suffer from uncertainty and, more importantly, the model would not generalize to other spatial regions. Thus, spatial features are usually derived from the context of the spatial location, here public POI data, since the categories of surrounding POIs are a valuable predictor~\citep{yan2017itdl} for the user's location category. 
POI data are, for example, available from the public Foursquare API or from Open Street Map (OSM). In either case, the dataset includes geographic point data and a categorization taxonomy of broad and more specific POI labels, 
e.g., a POI may be part of both the ''Shoe Store'' and the overarching ''Retail'' category. For most spatial features, we only use the broadest level and denote its categories as $\Psi = \{\psi_1, \dots, \psi_n\}$. A POI $p$ has a set of coordinates $(x(p), y(p))$, and is assigned to a main POI category, $c_p(p)$.\footnote{Note that our notation explicitly distinguishes \textit{location} categories ($c(l_i^u)\in C$) from \textit{POI} categories ($c_p(p) \in \Psi$), since they may be different. For example, an attacker could use POI data with 10 categories ($|\Psi|=10$) to classify user location data into only three categories such as $C=\{\text{Work, Leisure, Eating}\}$.} For example, $p$ may be assigned to $c_p(p) = \psi_2 = \text{Retail}$.

In the literature, different approaches have been used to extract features from the POI distribution around a specific point. We found empirically that a combination of the following methods yields the best results for our attacker's task:
\begin{itemize}
    \item \textbf{Category-count of the k-nearest POIs}: Given a location $(x(l_i^u), y(l_i^u))$, the $k$ closest POIs $p_1,\dots, p_k$ are found via a ball tree search, and the count of each category among those is computed. The result is a feature vector where the first element corresponds to the number of occurrences of the \textit{first} category among the $k$ closest POIs and accordingly for the other categories; formally 
    \begin{gather}
    \Big[
    \sum_{i=1}^k \mathbbm{1}[c_p(p_i)=\psi_1],\ \  
    \sum_{i=1}^k \mathbbm{1}[c_p(p_i)=\psi_2], \ \ 
    \dots
    \Big]
    \end{gather}
    Furthermore, as an indicator of the POI density at $(x,y)$, the mean distance from the $k$ nearest POIs is extracted as a feature. We set $k=20$ in our experiments.
    \item \textbf{Count and distance of POIs within a fixed radius}: The semantic attack requires more specific distance information of the POIs for each category. For example, if there is no restaurant within 1km, it is unlikely that the location category is ''Dining''. Thus, we consider all POIs around $(x(l_i^u), y(l_i^u))$ within a specified radius $r$, denoted as the set $P(x, y, r)$\footnote{for brevity, we omit $l_i^u$ here}, and again compute the count of each category. 
    \begin{gather}
    \Big[
    \sum_{p\in P(x, y,r)} \mathbbm{1}[c_p(p)=\psi_1],\ \  
    \sum_{p\in P(x, y,r)} \mathbbm{1}[c_p(p)=\psi_2], \ \ 
    \dots
    \Big]
    \end{gather}
    In addition, we consider the minimum distance of POIs of one category to the location:
    \begin{gather}
    \Big[
    \min_{\substack{p\in P(x, y,r) \\ c_p(p) = \psi_1}} \Vert \big(\substack{x \\ y}\big) - \big(\substack{x (p) \\ y(p)}\big) \Vert,\ \  
    \min_{\substack{p\in P(x, y,r) \\ c_p(p) = \psi_2}} \Vert \big(\substack{x \\ y}\big) - \big(\substack{x (p) \\ y(p)}\big) \Vert, \ \ 
    \dots
    \Big] 
    \end{gather}
    We set the radius to $200$m based on the results of preliminary experiments. If a category does not appear within the radius, we fill the corresponding vector field by the radius $r$. As an example, consider that three POIs are found within radius $r=200$m of the location: $p_1$ of category $\psi_3$ with 50m distance, $p_2$ of category $\psi_2$ with 10m distance, and $p_3$ of category $\psi_2$ with 80m distance. The resulting vectors (assuming there are only three categories) are $[0, 2, 1]$ and $[200, 10, 50]$.
    \item \textbf{Space2vec:} 
    In contrast to hand-crafted features based on distance and category counts, there is the option to \textit{learn} coordinate representations. 
    The task of finding an efficient and informative representation of points, dependent on their coordinates and POI context, was tackled recently in work on space embeddings. We employ the state-of-the-art \textit{space-to-vec} approach by \cite{mai2020multi}. Inspired by word embeddings in natural language processing, the idea is to learn a compact vector representation for points. The training is based on a supervised learning task, namely to distinguish surrounding points from unrelated, arbitrary distant samples that were drawn as negative samples. We deploy their public code base\footnote{\url{https://github.com/gengchenmai/space2vec}} to train the algorithm on our POI datasets $\mathcal{P}$, including the first \textit{two} category levels. 
    Specifically, we split $\mathcal{P}$ into training, validation (10\%), and testing set (10\%) and employ the \textit{joined} approach by \cite{mai2020multi}; i.e., training a location decoder and a spatial context decoder jointly. We set the embedding size to 16 but retained all other parameters as suggested by the authors. The model, which was trained only on $\mathcal{P}$, can be applied on a new location given its coordinates and its spatial context (coordinates and categories of the surrounding POIs) as input.
\end{itemize}

\subsection*{Model training}
\subsubsection*{Machine learning model}\label{sec:xgboost}

We chose the XGB approach over other machine learning models for its interpretability and its suitability for unbalanced data, rendering it superior in many applications. Nevertheless, we also implemented a multi-layer perceptron (MLP) for comparison. A model was implemented with two layers of 128 neurons respectively, with dropout regularization, ReLU activation and a softmax function in the output layer. The network was trained with the AdamOptimizer (learning rate 0.001) and with early stopping. For the XGBoost model, we utilize the XGBoost implementation in the \texttt{xgboost} Python package\footnote{\url{https://xgboost.readthedocs.io/en/stable/python/python_intro.html}} and only tune the parameter that determines the maximum depth of the base learners. A depth of 10 turned out most suitable in our experiments. 
The MLP also exhibits good place categorization ability, but was consistently inferior to XGB. For example, with the Foursquare data for NYC and an obfuscation radius of 100m, the accuracy is 52.2\% for the MLP compared to 59.4\% for XGB (41.6\% vs 49.8\% for 200m obfuscation, etc.). We, therefore, only report the results for XGB in this study.

\subsubsection*{Location masking}

A simple protection method for the use of location-based services is a random displacement of the coordinates to mask the real location. For example, iPhone users can withhold the precise locations from applications and only allow them to access the ''approximate'' location.
Here, we utilize location obfuscation to model imprecise GNSS data or basic data protection. The user's location is simply replaced by a new location sampled from a uniform distribution within a given radius $r$ (see \autoref{fig:main}a). 
Note that we focus on the obfuscation of the spatial information and leave the possibility of masking temporal information as in~\citep{mckenzie2016geo} for future work on semantic privacy. 
After the location masking step  (\autoref{fig:main}a), the raw (and obfuscated) spatio-temporal data are featurized (\autoref{fig:main}b) by deriving temporal features from the check-in time and spatial features from the coordinates matched with public POI data.

\subsubsection*{Data split}

We test for the attacker's accuracy by splitting the data into train and test sets, as shown in \autoref{fig:main}c. 
By default, the dataset is split by user, 
i.e., 10\% of the users are taken as the test set while the model is trained on 90\%. In practice, we report all results upon \textit{10-fold cross validation} such that all users were part of the test set once. The results simulate the scenario where the attacker obtains a labeled train dataset from a specific region and utilizes it to train an ML model with the goal to infer location profiles of new users but in the same region. However, the attacker may not always have labeled data from exactly the same spatial region. To analyze this scenario, we additionally simulate the attack with a \textit{spatial split}. In detail, the dataset is divided by separating the x- and y- coordinates in a $3\times 3$ grid, to yield nine roughly equal-sized subsets. The samples from each grid cell are used as the test set once.



\subsection*{Availability of data and materials}
All source code for reproducing our results is published on GitHub: \url{https://github.com/mie-lab/trip_purpose_privacy}.
The Foursquare data is publicly available at \url{https://sites.google.com/site/yangdingqi/home/foursquare-dataset?pli=1}. 

\subsection*{Competing interests}
The authors declare that they have no competing interests.


\subsection*{Authors' contributions}
N.W., O.K. and K.J. conceptualized the project. N.W. and O.K. performed 
the literature research. N.W. developed the methodology, implemented the
 algorithms, prepared all visualizations and wrote the main manuscript 
draft. K.J. revised the manuscript. O.K. and M.R. supervised the project
 and reviewed the manuscript.

\newpage

\bibliographystyle{abbrvnat}
\bibliography{references}





\section*{Appendix}
\input{appendix}

\end{document}

%% file: appendix.tex
\subsection*{Scenario comparison results}

\autoref{tab:results_100} provides an overview of the considered scenarios and metrics, and lists the results for protected location data with an obfuscation radius of 100m as an example. In the first scenario, both the check-in train data and the POI data are taken from the Foursquare dataset, and a machine-learning attack with the XGBoost model trained on spatio-temporal features can achieve a place categorization accuracy of 61.6\%. This is remarkable considering that there are 12 categories. 
It further translates to a user profiling error\footnote{The user profiling error is the average Euclidean distance between real and predicted profiles, $E_{\hat{p}(u), p(u)}$} of 0.124 which corresponds to a top-5 user re-identification accuracy (hit@5) of 40.4\%. In other words, given (obfuscated) location data and temporal visitation patterns, the attacker can infer an approximate profile $\hat{p}(u)$ for user $u$ such that, in 40.4\% cases, his real profile $p(u)$ is among the five real user profiles most similar to $\hat{p}(u)$.

\input{figures/results_100_table.tex}

\begin{figure}[htb]
     \includegraphics[width=\columnwidth]{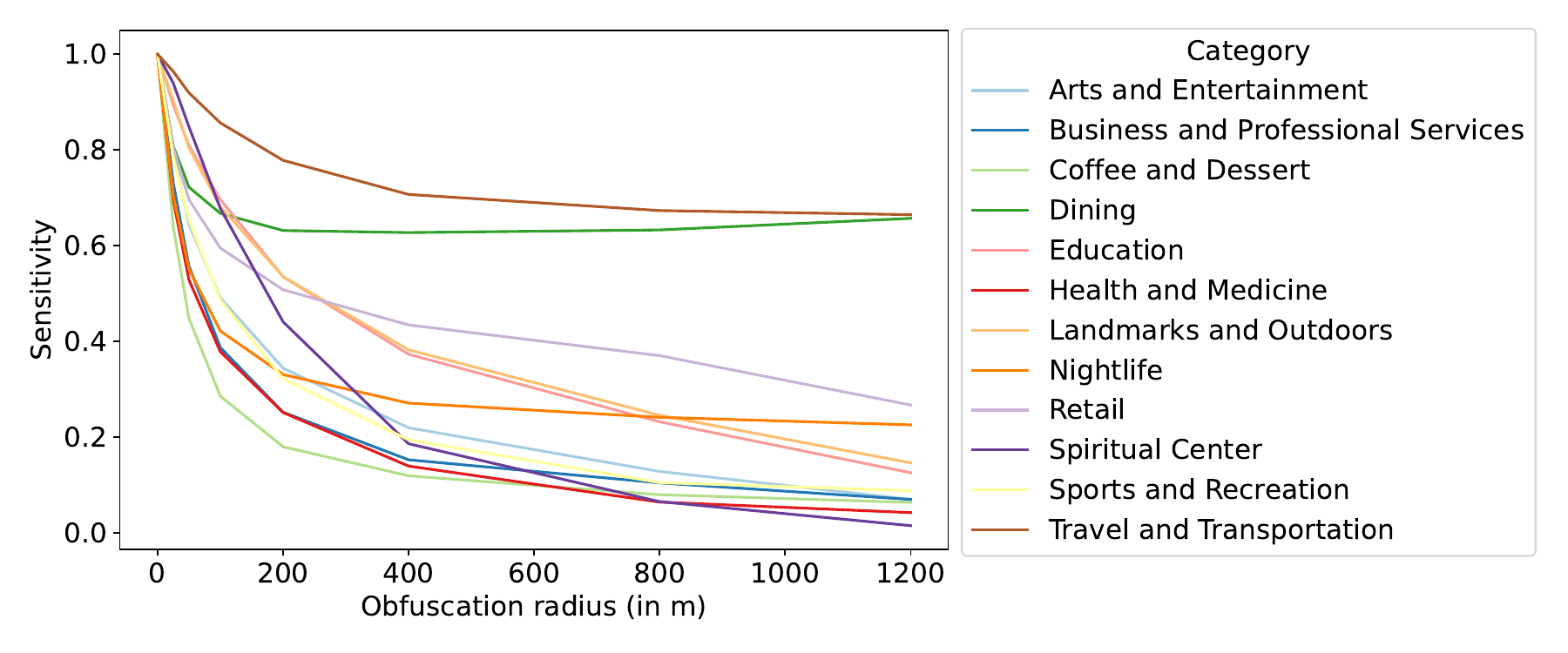}
    \caption{Sensitivity of place recognition by category}
    \label{fig:sensitivity_category}
\end{figure}

Furthermore, as shown shown in the confusion matrix in \autoref{fig:confusion}, the place classification accuracy clearly differs between categories. A more detailed analysis of this effect is provided in \autoref{fig:sensitivity_category}. At stronger obfuscation, many places are 
erroneously labeled as ''Dining'' or ''Travel and Transportation''. Therefore, the sensitivity for these places remains high, while decreasing for the other place categories.

\FloatBarrier

\subsection*{Tokyo vs NYC - effect of location obfuscation}

\begin{figure}[htb]
\begin{minipage}{.48\textwidth}
     \includegraphics[width=\columnwidth]{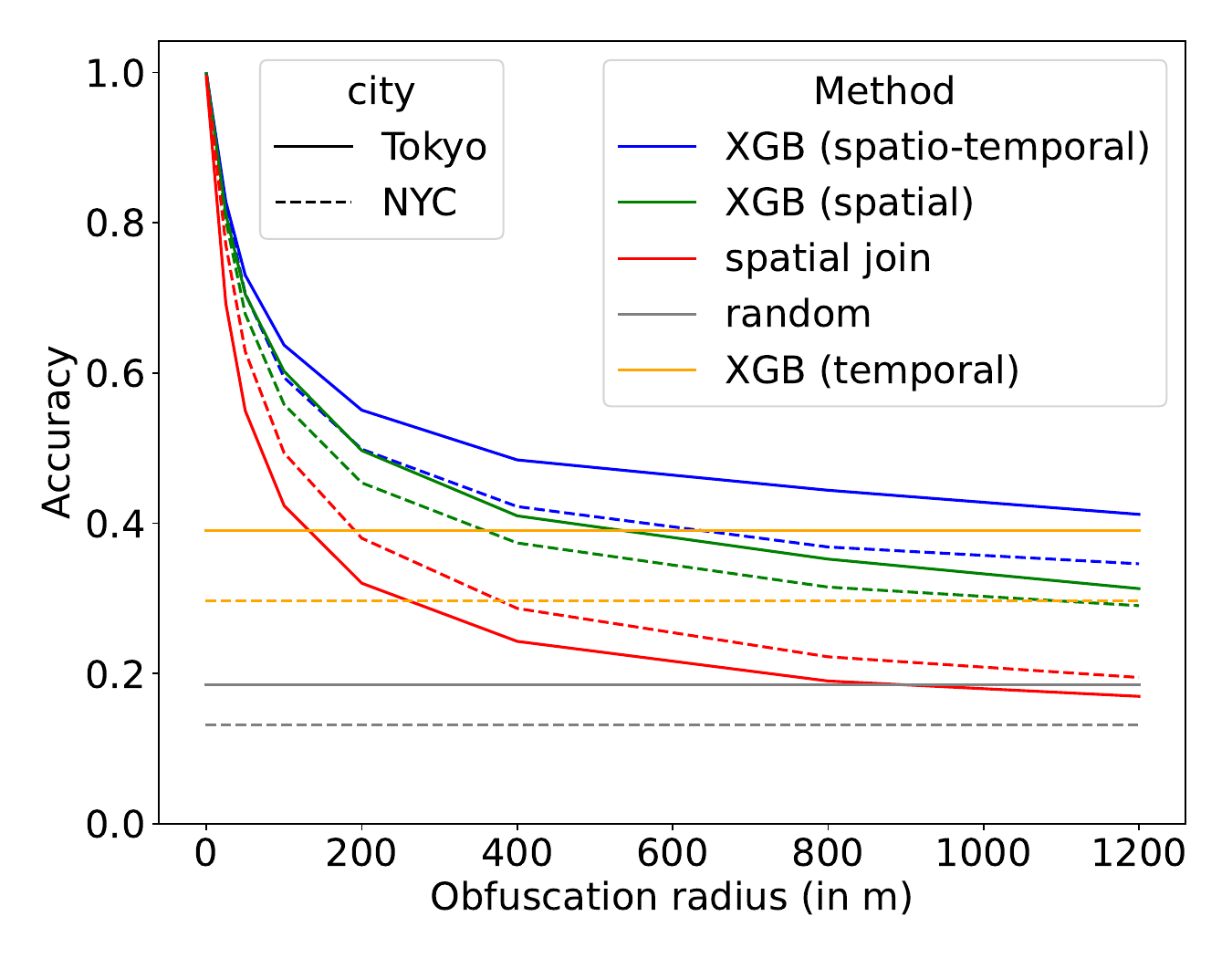}
    \caption{Place categorization accuracy of Tokyo compared to NYC}
    \label{fig:ny_tokyo} 
\end{minipage}
\hfill
\begin{minipage}{.48\textwidth}
\includegraphics[width=\columnwidth]{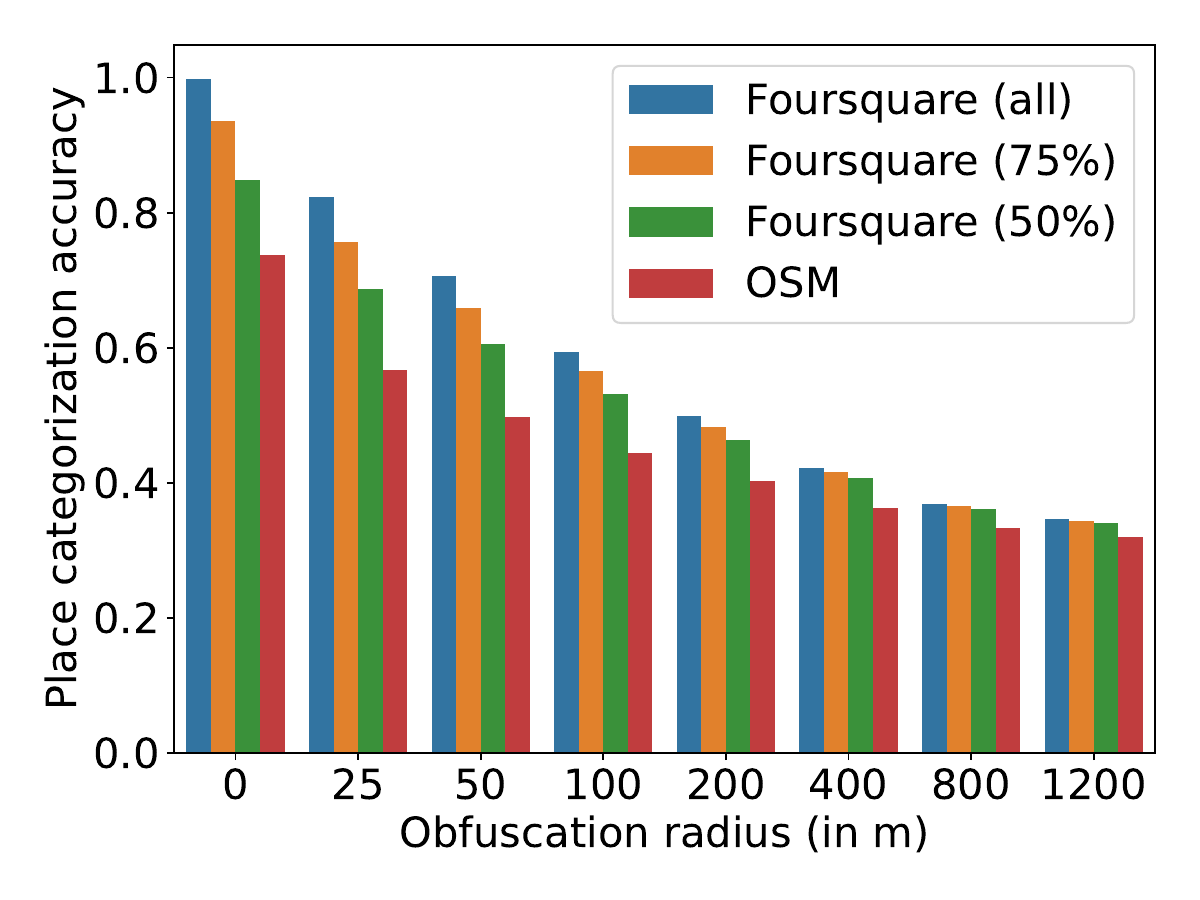} 
\caption{Effect of reduced POI context data quality on task 1 (place categorization)}
\label{fig:poi_user}
\end{minipage}

\end{figure}

In \autoref{fig:ny_tokyo}, we compare the results on two diverse cities, NYC and Tokyo. While the results show the same decrease- and convergence behavior in both cities, the accuracy is generally larger for Tokyo. This is, however, due to label imbalance in Tokyo which leads to a better performance of the random baseline (grey lines \autoref{fig:ny_tokyo}). 

\FloatBarrier

\subsection*{Effect of POI data completeness on place categorization task}

\autoref{fig:poi_user} provides the results for the first task (place categorization) by POI data completeness, complementing the user-identification results in \autoref{fig:poi_quality}. As was already seen for user-identification, the accuracy decreases disproportionally.


\FloatBarrier

\subsection*{User profiling error}

\begin{figure}[t]
\begin{minipage}{.48\textwidth}
    \includegraphics[width=\columnwidth]{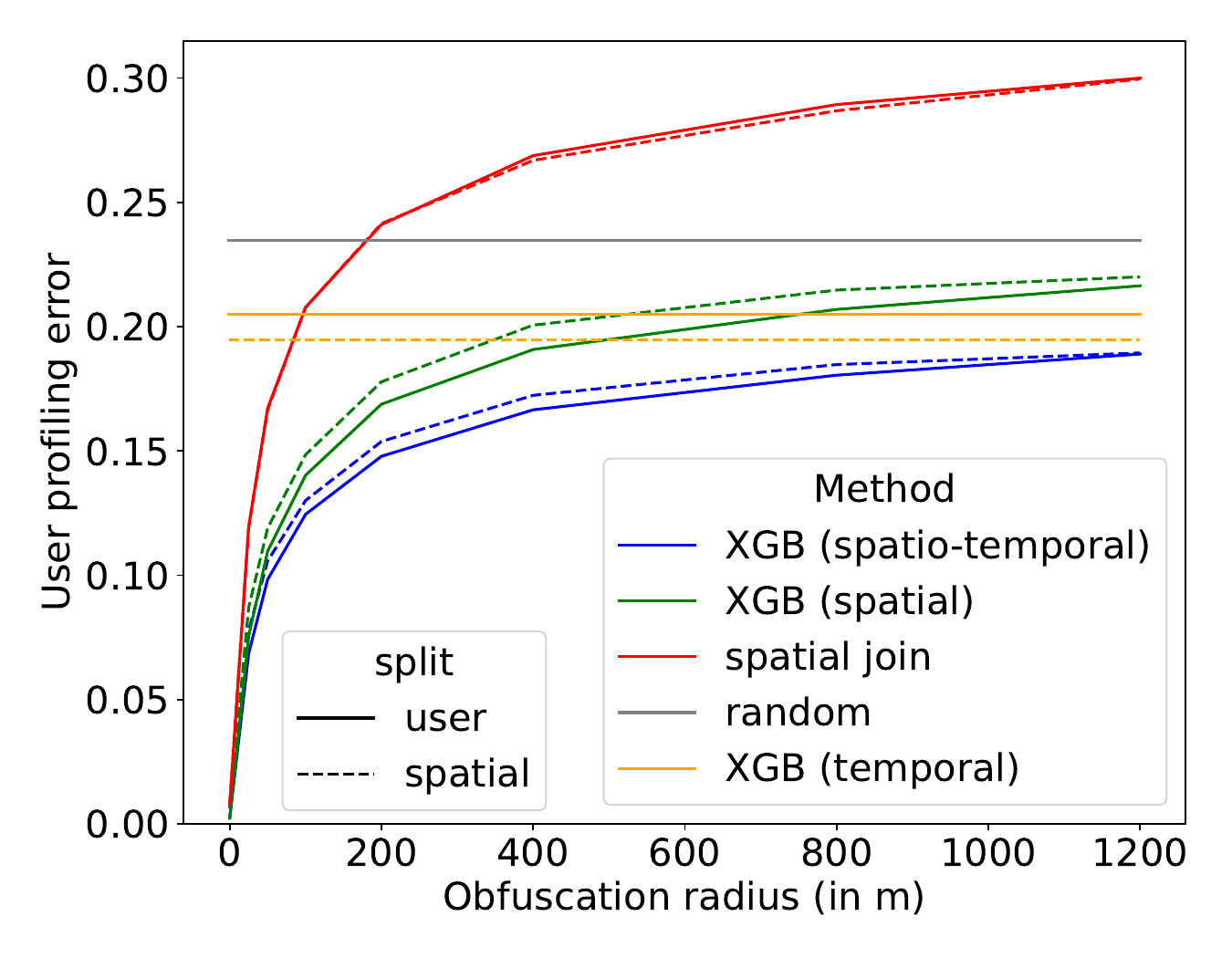}
    \caption{User profiling error (Euclidean distance between true and predicted user profile) by obfuscation radius}
    \label{fig:user_mae}
\end{minipage}
\hfill
\begin{minipage}{.48\textwidth}
    \centering
    \includegraphics[width=\textwidth]{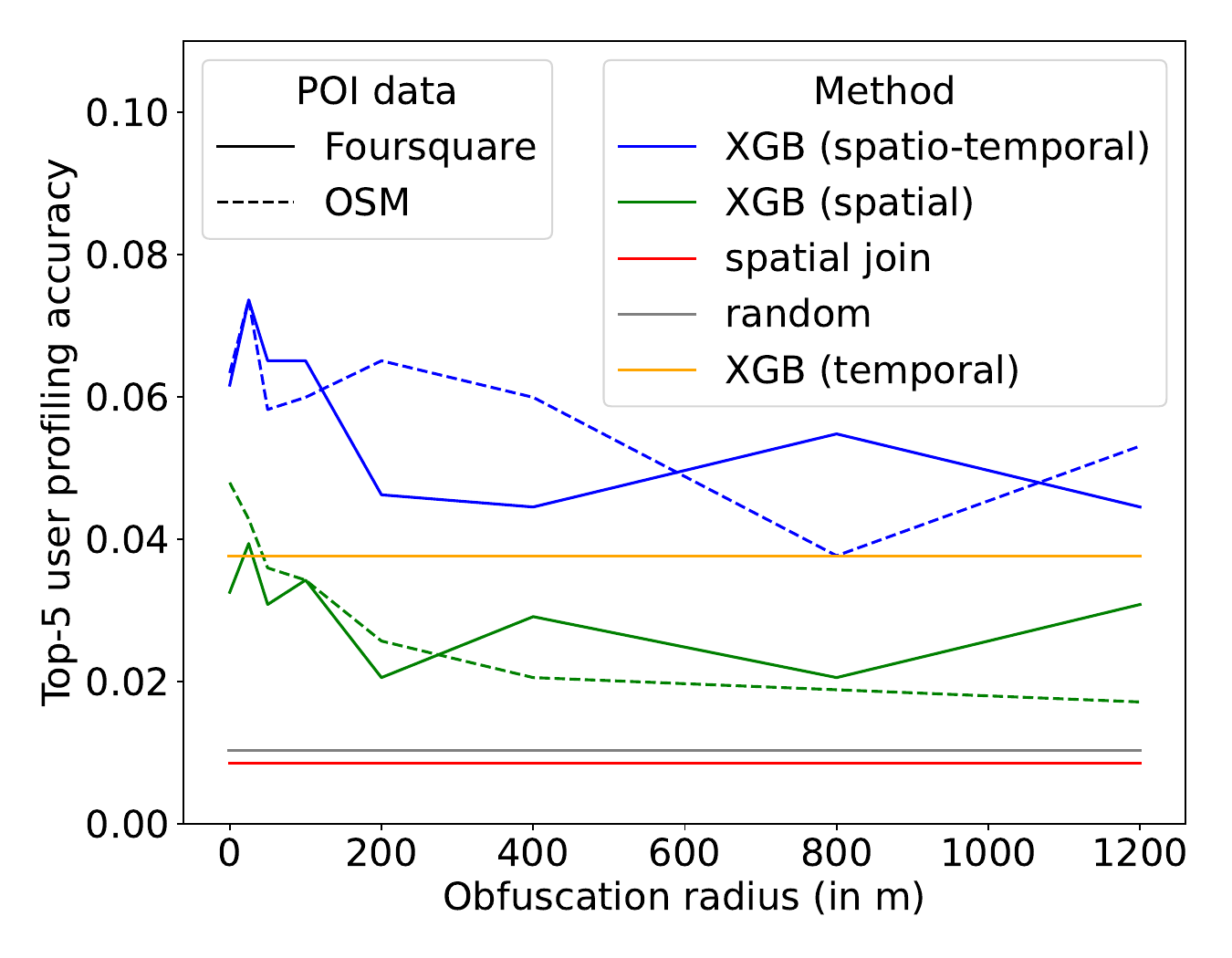} 
    \caption{Profiling results on GNSS tracking data. The profiling accuracy is low even with low location obfuscation, and the relation between obfuscation radius and profiling accuracy is noisy.}
    \label{fig:yumuv}
\end{minipage}
\end{figure}

Corresponding to the user identification accuracy presented in \autoref{fig:user_profiling_results}, we show the changes of the profiling errors in \autoref{fig:user_mae}, where the error is the Euclidean distance between the user's real profile vector $p_c(u)$ and the estimated profile vector $\hat{p}_c(u)$.

\subsection*{Study on GNSS tracking data}\label{sec:tracking_data}

The Foursquare check-in dataset is suitable for our analysis since it is reliably labeled with detailed place categories. For comparison, we evaluate results on a GNSS-based tracking dataset from the so-called yumuv study. yumuv is a tracking study in Switzerland that was conducted to investigate the mobility behavior with a micro-mobility-bundle subscription~\citep{reck2022mode}. Participants of the study were tracked via the Myway app\footnote{\url{https://www.sbb.ch/en/timetable/mobile-apps/myway.html}} for 2-3 months and manually labeled their activities. The app provides a preprocessed version of the raw GNSS data, namely (labeled) staypoints and triplegs. We further use the Python library Trackintel~\citep{martin2023trackintel} to group staypoints into \textit{locations} via DBSCAN. As for Foursquare, we remove the ''home'' check-ins as well as further categories that are uninformative, such as ''wait'', ''errand'', and ''unknown''. This leads to a set of five location categories, namely C = \{'Work', 'Education', 'Sports and Recreation', 'Dining', 'Leisure'\}. The categorical distribution is shown in \autoref{fig:checkins_yumuv}.


The dataset is more challenging due to noisy GNSS tracking data as well as unreliable and incomplete activity labels (the labels were estimated by an app and manually corrected by the user). Additionally, different users hardly visit the same places, and, thus, a training dataset has limited value for an attacker. Indeed, the main source of information for the attacker are temporal features in the case of the yumuv data. \autoref{fig:yumuv} shows that the attacker can still achieve semantic inference that is significantly better than random; however, spatial information only plays a limited role. Consequently, location protection hardly hinders the semantic attack, and the performance converges already at an obfuscation radius of 100m. In \autoref{tab:results_100}, other metrics are listed, and it is also shown that the danger for privacy on a user level is limited, with a top-5 user identification accuracy of only up to 7.7\% (100m location obfuscation). We hypothesize that the main reason for the limited value of spatial context data for the attacker lies in the activity categories of the yumuv study (see \autoref{fig:checkins_yumuv}), including ''Work'' and ''Leisure'' as the largest classes, which are too generic and too difficult to locate with POI data. 


\FloatBarrier

\subsection*{Dataset analysis}

\begin{figure}[htb]
    \centering
    \begin{subfigure}[b]{0.47\columnwidth}
    \includegraphics[width=\textwidth]{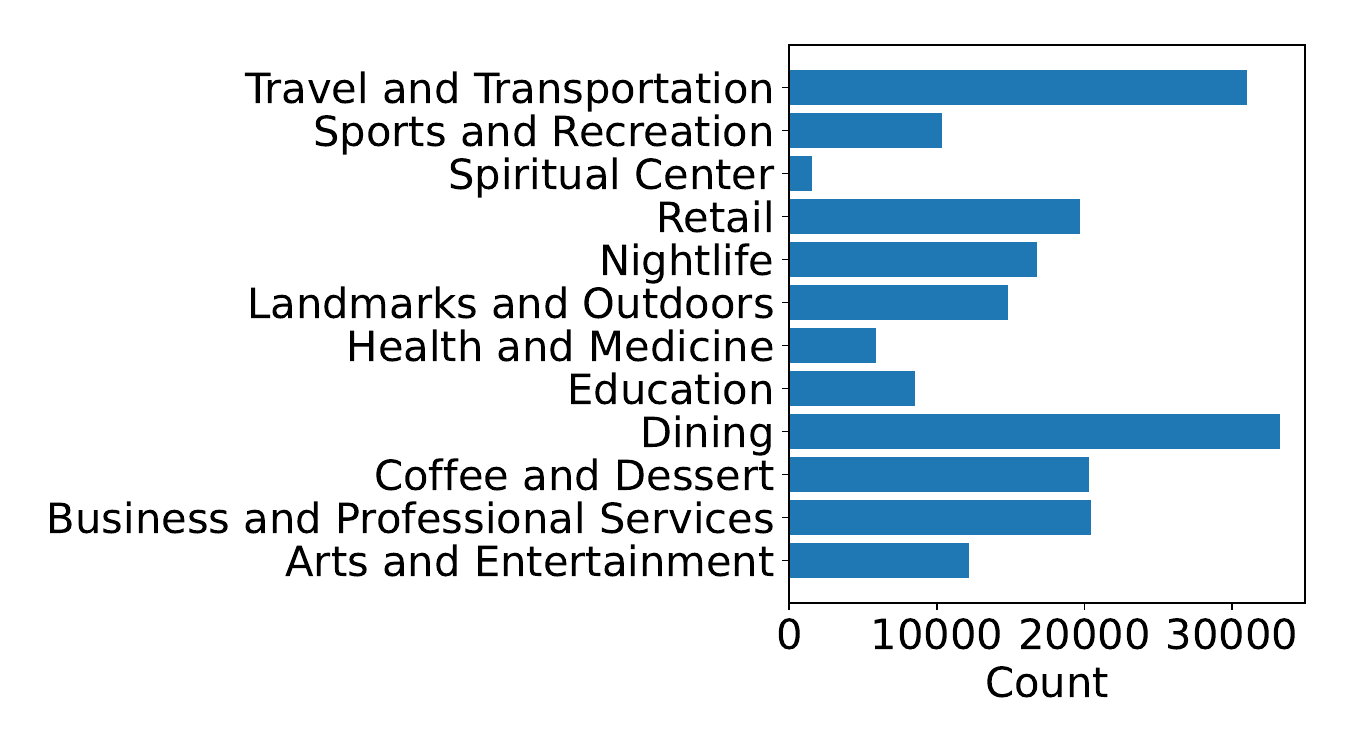}
    \caption{Foursquare check-ins}
    \label{fig:checkins_foursquare}
    \end{subfigure}
    \begin{subfigure}[b]{0.47\columnwidth}
    \includegraphics[width=\textwidth]{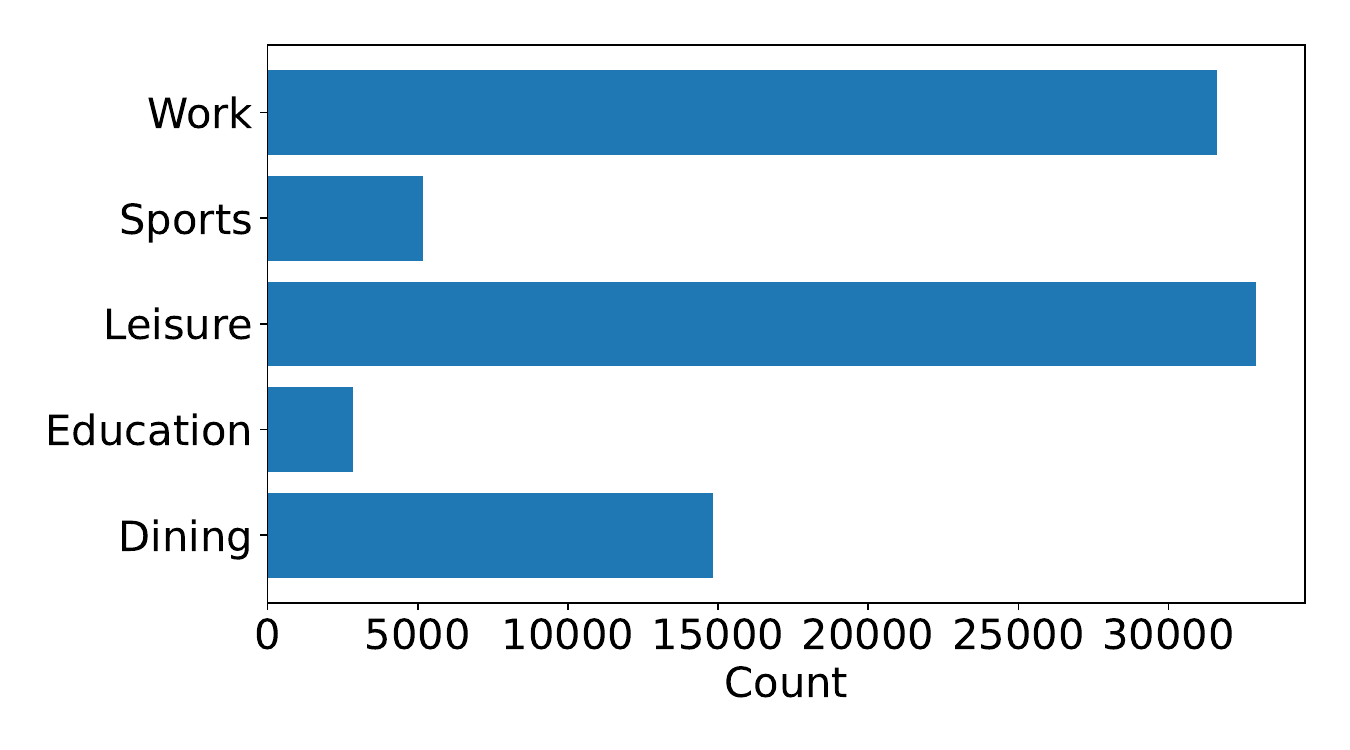}
    \caption{Yumuv tracking data}
    \label{fig:checkins_yumuv}
    \end{subfigure}
    \caption{Location category counts in Foursquare and Yumuv datasets}
    \label{fig:checkins}
\end{figure}

In \autoref{fig:checkins}, the distribution of category types over visited places is shown for Foursquare (\autoref{fig:checkins_foursquare}) and the yumuv (\autoref{fig:checkins_yumuv}) data.


\FloatBarrier

\subsection*{Analyzing place category autocorrelation in a semivariogram}\label{appendix_a}

The categories of user locations are predicted based on information from surrounding public POIs. Even if the location is obfuscated, the category of the closest POI is a good predictor for the location’s category. This is due to the spatial autocorrelation of place categories. For example, there are urban areas with many restaurants in one street, such that there is a high probability of finding several nearby places with the same category. We relate our results to the spatial autocorrelation of POIs by comparing our place categorization accuracy (\autoref{fig:main_result}) to the semivariogram of place category differences. To compute the variogram, we bin the distances and report the category correspondence in each bin, thereby accounting for the categorical nature of place categories and the continuous nature of the place coordinates. In detail, given a minimum and maximum distance $d_{min}, d_{max}$, the category variance is computed as 
$$ \gamma(d_{min}, d_{max})  = \frac{\sum_{l_1, l_2 \in N(d_{min}, d_{max})} \mathbbm{1}[c(l_1) == c(l_2)]}{|N(d_{min}, d_{max})|}  $$ 
where $N(d_{min}, d_{max})$ is the set of all pairs of locations with $d_{min} < d(l_1, l_2) \leq d_{max}$, and $\mathbbm{1}$ is the indicator function that counts the number of pairs with corresponding category.
Due to the high number of POIs in the dataset, we only take a 20 x 20km subregion of New York City and sample pairs of POIs randomly. 

\begin{figure}[htb]
    \centering
    \includegraphics[width=0.6\columnwidth]{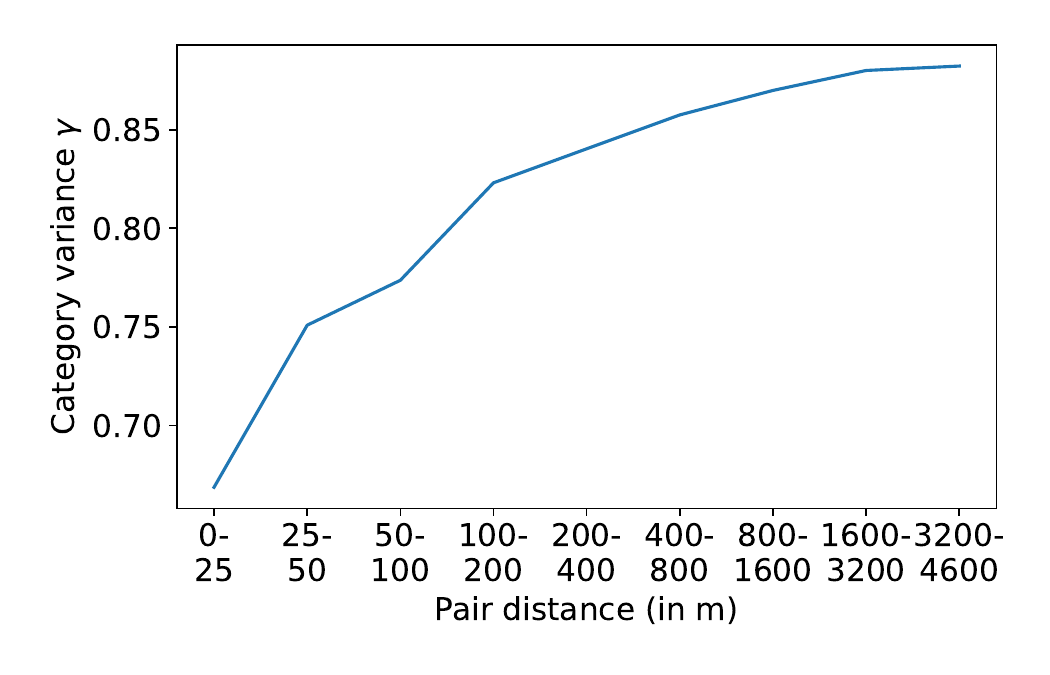}
    \caption{Semivariogram for measuring spatial autocorrelation of place categories. The y-axis shows the percentage of POI-pairs with different categories.}
    \label{fig:variogram}
\end{figure}

\autoref{fig:variogram} shows the change of $\gamma$ with increasing $d_{min}$ and $d_{max}$, as common in a semivariogram. Even places with less than 25m distance only have a 33\% chance of their categories to correspond ($\gamma(0, 25) = 0.67$). At a distance of more than 3.2km, the category variance converges to around 89\%, which is slightly lower than the expected variance of 8.3\% for 12 categories. However, the imbalance of categories explains this difference. The semivariogram confirms our finding in \autoref{fig:main_result} that spatial context data is hardly informative for location data that is obfuscated more than 1km.

%% file: figures/results_100_table.tex
\begin{table*}[htb]
    \centering
    \resizebox{\textwidth}{!}{

\begin{tabular}{l|l|l|c|ccc}
\toprule
& & & \textbf{Place categorization} & \multicolumn{3}{c}{\textbf{User profiling}}\\
\textbf{Check-in data }& \textbf{POI data} & \textbf{Method}           &  Accuracy &  Profiling error &
\begin{tabular}{c}
     Top-5 identifi- \\
     cation accuracy 
\end{tabular}
&  Privacy loss (median) 
                          \\
                          \midrule
Foursquare & Foursquare & random &     0.159 &                       0.235 &                          0.003 &                  1.000 \\
         (NYC \& Tokyo)                 &     & XGB (temporal) &     0.344 &                       0.205 &                          0.023 &                  1.112 \\
                          &     & spatial join &     0.459 &                       0.208 &                          0.183 &                  2.296 \\
                          &     & XGB (spatial) &     0.580 &                       0.140 &                          0.328 &                  5.574 \\
                          &     & XGB (spatio-temporal) &     0.616 &                       0.124 &                          0.404 &                 11.001 \\
                          & OSM & random &     0.157 &                       0.235 &                          0.002 &                  1.000 \\
                          &     & spatial join &     0.224 &                       0.341 &                          0.013 &                  1.051 \\
                          &     & XGB (temporal) &     0.328 &                       0.206 &                          0.022 &                  1.116 \\
                          &     & XGB (spatial) &     0.476 &                       0.172 &                          0.117 &                  1.949 \\
                          &     & XGB (spatio-temporal) &     0.503 &                       0.157 &                          0.160 &                  2.316 \\
\midrule 
Yumuv study  & Foursquare & random &     0.370 &                       0.367 &                          0.010 &                  1.000 \\
     (Switzerland)                     &     & spatial join &     0.453 &                       0.698 &                          0.009 &                  1.000 \\
                          &     & XGB (spatial) &     0.541 &                       0.331 &                          0.034 &                  1.095 \\
                          &     & XGB (temporal) &     0.553 &                       0.333 &                          0.038 &                  1.108 \\
                          &     & XGB (spatio-temporal) &     0.618 &                       0.293 &                          0.065 &                  1.225 \\
                          & OSM & random &     0.372 &                       0.367 &                          0.010 &                  1.000 \\
                          &     & spatial join &     0.384 &                       0.698 &                          0.009 &                  1.000 \\
                          &     & XGB (spatial) &     0.541 &                       0.332 &                          0.034 &                  1.093 \\
                          &     & XGB (temporal) &     0.553 &                       0.333 &                          0.038 &                  1.108 \\
                          &     & XGB (spatio-temporal) &     0.623 &                       0.289 &                          0.060 &                  1.256 \\
\bottomrule
\end{tabular}

}
    \caption{Comparing attack-scenarios with an obfuscation radius of $r=100$. For the place categorization task, the accuracy over location-visit events is reported. User-profiling error, user identification accuracy and privacy loss measure the success of the attacker in user-profiling. 
    }
    \label{tab:results_100}
\end{table*}